\documentclass[10pt]{elsarticle} 
\usepackage{booktabs} 
\usepackage{tabularx}
\usepackage[unicode]{hyperref}

\journal{Journal of Systems and Software}

\usepackage{graphicx}
\usepackage{svg}
\usepackage{amsmath}
\usepackage{todonotes}
\usepackage{natbib}
\usepackage{textcomp}
\usepackage{amstext}
\usepackage{amssymb}
\usepackage{amsmath}
\usepackage{multicol}
\usepackage{tikz}
\usepackage{lettrine}
\allowdisplaybreaks
\usepackage[export]{adjustbox}
\usepackage{ragged2e}
\usepackage{subfig}
\usepackage{float}
\usepackage{booktabs}
\usepackage{makecell, multirow}
\usepackage{tabularx}
\usepackage{siunitx}
\usepackage{here}
\usepackage{longtable}

\newcolumntype{C}[1]{>{\centering\arraybackslash}m{#1}}
\usepackage{url}
\newcounter{finding}
\newenvironment{finding}[1][]{\refstepcounter{finding} \par\medskip\hrule\par\medskip\small
   \noindent \textbf{Finding~\thefinding. #1} \rmfamily}{\medskip\hrule\medskip}

\begin{document}

\begin{frontmatter}

\title{Prevalence, Common Causes and Effects of Technical Debt: Results from a Family of Surveys with the IT Industry}

\author{Robert Ramač}
\address{University of Novi Sad, Serbia}
\ead{ramac.robert@uns.ac.rs}

\author{Vladimir Mandić\corref{vladimirmandic}}
\address{University of Novi Sad, Serbia}
\ead{vladman@uns.ac.rs}

\author{Nebojša Taušan}
\address{INFORA Research Group, Serbia} 
\ead{nebojsa.tausan@infora.rs}

\author{Nicolli Rios}
\address{Federal University of Rio de Janeiro, Brazil} 
\ead{nicolli@cos.ufrj.br}

\author{Sávio Freire}
\address{Federal University of Bahia and Federal Institute of Ceará, Brazil} 
\ead{savio.freire@ifce.edu.br}

\author{Boris Pérez}
\address{Universidad de Los Andes and Francisco de Paula Stder. Univ, Colombia} 
\ead{borisperezg@ufps.edu.co}

\author{Camilo Castellanos}
\address{Universidad de Los Andes, Colombia} 
\ead{cc.castellanos87@uniandes.edu.co}

\author{Darío Correal}
\address{Universidad de Los Andes, Colombia} 
\ead{dcorreal@uniandes.edu.co}

\author{Alexia Pacheco}
\address{University of Costa Rica, Costa Rica} 
\ead{alexia.pacheco@ucr.ac.cr}

\author{Gustavo Lopez}
\address{University of Costa Rica, Costa Rica} 
\ead{gustavo.lopez_h@ucr.ac.cr}

\author{Clemente Izurieta}
\address{Montana State University and Idaho National Laboratories, United States} 
\ead{clemente.izurieta@montana.edu}

\author{Carolyn Seaman}
\address{University of Maryland Baltimore County, United States} 
\ead{cseaman@umbc.edu}

\author{Rodrigo Spinola}
\address{Salvador University and State University of Bahia, Brazil} 
\ead{rodrigo.spinola@unifacs.br}



\cortext[vladimirmandic]{Corresponding author}


\begin{abstract}
\textit{Context:}
The technical debt (TD) metaphor describes
actions made during various stages of software development that lead to a more costly future regarding system maintenance and evolution. 
According to recent studies, on average $25\%$ of development effort is spent, i.e. wasted, on TD caused issues in software development organizations. However, further research is needed to investigate the relations between various software development activities and TD.

\noindent\textit{Objective:}
The objective of this study is twofold. First, to get empirical insight on the understanding and the use of the TD concept in the IT industry. Second, to contribute towards precise conceptualization of the TD concept through analysis of causes and effects.

\noindent\textit{Method:}
In order to address the research objective a \textit{family of surveys} was designed as a part of an international initiative that congregates researchers from 12 countries---InsighTD.
At country level, national teams ran survey replications with industry practitioners from the respective countries. 

\noindent\textit{Results:}
In total 653 valid responses were collected from 6 countries.
Regarding the prevalence of the TD concept 22\% of practitioners have only theoretical knowledge about it, and 47\% have some practical experiences with TD identification or management. Further analysis indicated that senior practitioners who work in larger organizations, larger teams, and on larger systems are more likely to be experienced with TD management. 
Time pressure or \textit{deadline }was the single most cited cause of TD. Regarding the effects of TD: \textit{delivery delay}, \textit{low maintainability}, and \textit{rework} were the most cited.

\noindent\textit{Conclusion:}
InsighTD is the first family of surveys on technical debt in software engineering. 
It provided a methodological framework that allowed multiple replication teams to conduct research activities and to contribute to a single dataset. Future work will focus on more specific aspects of TD management.  
\end{abstract}

\begin{keyword}
Technical Debt \sep Survey \sep InsighTD \sep Causes of technical debt \sep Effects of technical debt
\end{keyword}

\end{frontmatter}

\section{Introduction}
\label{sec.Indtroduction}

Software companies are often in situations that require them to make trade-offs between software quality and preset deadlines and goals. Making such trade-offs may result in short-term wins, but long-term it may introduce friction in software development and maintenance processes. This is aptly described with a metaphor of \textit{technical debt} \cite{dagstuhl2016managing,kruchten2012technical,shull2013technical,TD2018tertiaryStudy}. The concept of technical debt (TD) describes a phenomenon that impacts software projects and makes them difficult to manage \cite{TD2018tertiaryStudy}. For example, lack of coding standards can result with bad code, i.e. code debt and code smells \cite{TD2016identification,TD2018tertiaryStudy,zazworka2014comparing}.
It seems that TD issues are prevalent in organizations and teams of all sizes. 
Some recent studies indicated that on average $25\%$ of development effort is spent---wasted---on TD caused issues \cite{besker2019software,besker2018technical}.

The technical debt metaphor was coined by Ward Cunningham in the 90s with an explicit intention to better communicate important aspects of internal software quality---software attributes that are not directly visible to the users \cite{buschmann2011pay,TD2018tertiaryStudy, kruchten2012technical}. Interestingly in the early 2000s, software engineers were more occupied with the concept of value---software attributes that are directly visible to the users---and with approaches to deliver that value to the market more efficiently, e.g. Agile approaches, lean software development, Value-based Software Engineering \cite{beck2001manifesto, mandic2010flowing, boehm2003value}. Just in recent years, the concept of TD is gaining more attention from the research community as well as from the industry \cite{TD2018tertiaryStudy}.

Essentially, TD represents \textit{shortcuts} made during various stages of software development that result with a more costly future regarding system maintenance and evolution. A more elaborate \textit{definition} was formulated by McConnell \cite{mcconnell2008managing} as:
\textit{technical debt} contextualizes the problem of outstanding software development tasks 
as a kind of debt that brings a short-term benefit to the project
 that may have to be paid later in the development process with interest. For example, a poorly designed class tends to be more difficult and costly to maintain than if it had been implemented using good object-oriented practices.
Furthermore, McConnell distinguishes two kinds of technical debt: \textit{intentional}---purposefully injected by acknowledging benefits of doing so---and \textit{unintentional}---result of poor engineering practices \cite{mcconnell2008managing}.
Beside this definition, the literature recognizes many other definitions of TD \cite{dagstuhl2016managing,kruchten2012technical,shull2013technical,TD2018tertiaryStudy}. 

By definition, TD is a trade-off between short-term benefits and long-term effects, where taking short-term benefits causes some long-term issues, i.e. \textit{negative} consequences. However, there are some indications that in some situations injecting TD can have overall \textit{positive} effects. e.g. start-up companies \cite{TDStartUps}.

In past years, a significant effort was made by researchers to better understand and identify technical debt from the analysis of source code, i.e. code debt and code smells \cite{TD2016identification,TD2018tertiaryStudy,zazworka2014comparing}. However, this is just one perspective. Software development can be seen as a chain of various decisions (business, design, architecture, etc.), which starts well before coding, and all those decisions impact the creation of a product \cite{mandic2010flowing}, and consequently the creation of technical debt. 
One of the conclusions of the Dagstuhl Seminar was that all software development artifacts can carry technical debt \cite{dagstuhl2016managing}. Therefore, we need to better understand the connections between non-coding activities and TD,
such a cognition will allow us to balance a delivered value of software with accompanying technical debt.

The nature of the TD phenomenon itself calls for empirical approaches with sufficiently large and representative datasets. Such empirical studies are still rare.
However, there are empirical studies focused on understanding TD causes, effects, and practices to manage the accumulated debt \cite{martiniATD2017, martiniEUROMICRO2014,YLIHUUMO2016TD,YliHuumoPROFES2014,ernst2015measure}. For example, Yli-Huumo et al. conducted an interpretive case study with two mid-sized organizations in Finland, and they reported 7 scenarios of why workarounds are made in projects \cite{YLIHUUMO2016TD, YliHuumoPROFES2014}. Martini et al. investigated relations between software architecture and TD items \cite{martiniEUROMICRO2014}. In their case study, they involved 7 study units with 25 participants \cite{martiniEUROMICRO2014}. In 2015, Ernst et al. conducted a study---so far, the largest empirical investigation of TD---with 536 participants from 3 large organizations \cite{ernst2015measure}. The results of their study confirmed the usefulness of the metaphor, but at the same time they accentuated the lack of existing tools for managing TD.
The existing empirical studies are lacking the heterogeneity of the data sample, e.g. studies with large number of participants are conducted with a few organizations or in one country. Therefore, we need further empirical investigations that aim at increasing the representativeness of the sample and collected dataset.

The objective of this research is to collect representative empirical data about the TD concept, foremost about its prevalence within the IT industry alongside with the identification of TD causes and effects. This objective is further elaborated with the following research questions:
\begin{itemize}
\item[RQ1:] To what extent are software professionals familiar with the concept of TD? 
\item[RQ2:] What causes lead software development teams to incur TD? 
\item[RQ3:] What effects does TD have on software projects? 
\end{itemize}
In order to address the research objective a \textit{family of surveys} was designed as a part of an international initiative that congregates researchers from 12 countries---InsighTD.\footnote{\url{http://www.td-survey.com}}
This is the first aggregated data analysis of the \textit{InsighTD} survey. 
The empirical dataset consists of $653$ valid responses from industry practitioners in $6$ countries.
By publishing this report, we finalized the analysis of data for basic TD questions about the prevalence, causes, and effects. As a part of InsighTD research roadmap other aspects of TD management will be addressed in the future.\footnote{InsighTD research roadmap and publication map can be found on the link: \url{http://www.td-survey.com/publication-map/}}

Findings presented in this paper are of interest for industry practitioners as well as for researchers in the field of TD, who can benefit from the insights about the most common causes and effects of TD. 
The contributions of the paper are: 
\begin{enumerate}
\item Regarding the familiarity of the TD concept within the industry, 31\% of participants never heard about it, while 27\% of those who are familiar with it, have actively managed TD on projects. Furthermore, the only demographic variable that does not have significant effect on the TD familiarity is \textit{system age} ($\chi^2=15.192, p=0.2311$), while the other variables have a significant effect.    
\item Our findings show that various TD types are present in the industry. There is no single TD type that significantly stands out. We identified the top 5 TD types by occurrence as: \textit{design debt} (21.99\%), \textit{test debt} (19.94\%), \textit{code debt} (14.66\%), \textit{architecture debt} (10.7\%), and \textit{documentation debt} (9.09\%). 
\item Conceptualization of TD causes with a mind map, i.e. visualization of categories of causes and their frequencies. Identification of major TD causes. The single most cited TD cause is \textit{deadline}, i.e. time pressure caused by short deadline.
\item Conceptualization of TD effects with a mind map, i.e. visualization of categories of effects and their frequencies. Identification of major TD effects. The top three effects of TD are \textit{delivery delay}, \textit{low maintainability}, and \textit{rework}.
\end{enumerate}

The rest of the paper is organized as follows. After discussing the related work and our previous work in Section \ref{sec.RelatedWork}, we describe the research approach in Section \ref{sec.ResearchApproach}. Section \ref{sec.Results} presents the results of the survey. In Section \ref{sec.Discussion} we discuss findings and observations. Trustworthines of the study is discussed in Section \ref{sec.Trustworthiness}, followed by our final remarks in Section \ref{sec.Conclusion}.

\section{Related Work}
\label{sec.RelatedWork}

This section starts with the discussion about survey studies on TD (Section \ref{sec.SurveyResearchTD}). After this, Section \ref{sec.ResearchCausesEffects} describes the related work on causes and effects of TD. 
Our literature review process started with an initial dataset of references. The initial dataset was collected during our previous meta-study of systematic literature reviews on TD \cite{TD2018tertiaryStudy}. From the initial dataset we selected references that were focusing on empirical studies of TD causes and effects. The selected references were used as starting references for a form of the snowballing search.  
Two main inclusion criteria were: (1) survey-based research, i.e. empirical studies that used a survey as the main instrument, and (2) empirical studies focusing on TD causes and TD effects. 

Finally, in Section \ref{sec.PreviousWork} we discuss our previously disseminated results on the topic and how this work complements them.

\subsection{Survey Research on TD}
\label{sec.SurveyResearchTD}

Technical debt and related topics became a subject of empirical investigations after 2010. The early empirical investigations used qualitative, interview based, approaches \cite{taksande2011empirical}, because little was known about the phenomena. Later, this was followed by observational studies, experience reports, and case studies \cite{siebra2012managing,morgenthaler2012searching,tom2013exploration,zazworka2013case}. This early accumulation of knowledge about the TD phenomena paved a road for more structured empirical investigations using surveys. In this section, the extent and representativeness of the survey-based research on TD is presented.  

Table \ref{tab.SurveyResearchTD} presents surveys on TD excluding surveys and replications of the InsighTD project. Also, surveys that were used in case studies as pre-study or follow-up questionnaires are excluded. Since 2014, there were 11 conducted surveys. Regarding the sample size, it varies from 40 in \cite{silva2019taste} up to 536 in \cite{ernst2015measure}.
Research focus of the surveys mainly revolves around general TD topics, including, TD causes and effects, monitoring and management practices, productivity issues, and software quality (Table \ref{tab.SurveyResearchTD}). 

\begin{table}[h]
    \caption{Survey-based research on TD.}
    \label{tab.SurveyResearchTD}
    \centering \scriptsize
    \begin{tabular}{llm{5.3cm}llll} 
        \toprule
       & & & & \multicolumn{3}{c}{Representativeness} \\ 
        \cline{5-7}
        No. &\begin{tabular}[c]{@{}l@{}}Ref.\end{tabular} &  \multicolumn{1}{c}{Research Focus} &  \multicolumn{1}{c}{Year} & \multicolumn{1}{c}{Part.} & \multicolumn{1}{c}{Org.} & \multicolumn{1}{c}{Count.} \\ 
            \midrule
            1.&    \cite{TDandAgileSurvey2014} & General TD in Agile software development.    & 2014.    & 80  & -  & 3 \\
            2.&    \cite{ernst2015measure} & General TD causes.    & 2015.    & 536  & 3  & 1 \\
            3.&    \cite{rocha2017understanding} & TD management, focusing on the code quality and code debt.   & 2017.    & 74  & -  & 1 \\
            4.&    \cite{codabux2017empirical} & General TD and TD risk assessment.    & 2017.    & 67  & 33  & 14 \\
            5.&    \cite{Besker2017TDBill} & Estimation of TD costs.    & 2017.    & 258  & -  & - \\
            6.&    \cite{holvitie2018technical} & General TD in Agile software development.   & 2018.    & 184  & -  & 3 \\
            7.&    \cite{besker2018technical} & TD and software developer productivity.   & 2018.    & 43  & -  & - \\
            8.&    \cite{martini2018technical} & TD monitoring and tracking.   & 2018.    & 226  & 15  & - \\
            9.&    \cite{arvanitou2019monitoring} & TD monitoring.   & 2019.    & 60  & 11  & 9 \\              
            10.&    \cite{silva2019taste} & TD perception and TD management.    & 2019.    & 40  & 12  & - \\          
            11.&    \cite{apa2020taste} & General TD and TD management.   & 2020.    & 259  & -  & 1 \\ 
              \bottomrule
    \end{tabular}
\end{table}

The Ernst et al. \cite{ernst2015measure} survey is the largest with 536 valid answers, however the survey was conducted in 3 large organizations thus the results of the survey can be biased toward large organizations. On the other hand, Codabux et al. \cite{codabux2017empirical} managed to collect responses from 67 participants in 33 organizations resulting with a good diversity of company sizes. Furthermore, the study of Codabux et al. has international character, involving participants from 14 countries.

Overall, 3 studies are national \cite{ernst2015measure,rocha2017understanding,apa2020taste}, 4 studies are international \cite{TDandAgileSurvey2014,codabux2017empirical,holvitie2018technical,arvanitou2019monitoring}, and for the remaining 4 studies the country was not reported (Table \ref{tab.SurveyResearchTD}).  

In summary, although the survey as a research instrument is convenient for collecting responses from large samples, evidently survey research needs to deal with challenges of low response rates and consequently lower repressentativnes of the population. This was one of the main motivations to choose an international family of surveys for the research approach of this study.

\subsection{Research on Causes and Effects of TD}
\label{sec.ResearchCausesEffects}

Avgeriou et al. \cite{dagstuhl2016managing} indicated that the causes of TD can be a process, a decision, an action (or lack thereof), or an event that triggers the existence of a TD item, such as schedule pressure, unavailability of a key person, or lack of information about a technical feature. Also, they reported that TD can affect the value of the system, the costs of future changes, the schedule, and system quality. Those perceptions on causes and effects of TD reflect the consolidation of the opinion of a couple of researchers and practitioners during the Dagstuhl Seminar 16162. Nevertheless, there are other works on the topic, before and after the mentioned seminar, that support and, also, add new information to the TD causes and effects body of knowledge.

Codabux and Williams \cite{codabux2013managing} conducted an industrial case study to understand how TD affects the adoption of Agile software development. The study used participant observations, semi-structured interviews, and a questionnaire for collecting data. The authors analyzed the data using a coding scheme, where each question was related to a theme. For example, causes and impacts were associated with motivations behind incurring TD and consequences of TD, respectively. In total, 28 practitioners (product owners, scrum master, developers, and testers) participated in the case study. Regarding the causes of TD, the practitioners indicated that managerial decisions and resource constraints (a limited timeframe and the unavailability of developers to continue working on the feature) are the main causes for incurring TD. Starting the project from scratch was the main effect felt by practitioners. 

Yli-Huumo et al. \cite{YliHuumoPROFES2014} performed a case study of two product lines in a midsize Finnish software company for understanding the relationship between TD causes, effects, and management. The data was collected from interviews, and qualitative data analysis techniques were also applied. From 12 interviews, the authors found that the lack of time given for development, pressure put on the development team, complexity of the source code, business decisions (lack of technical knowledge and communication challenges), lack of coding standards and guides, junior coders, lack of knowledge about future changes, and lack of documentation are the causes of TD. On the other side, the effects recognized in the case study were time-to-market benefit, increased customer satisfaction, extra working hours, errors and bugs, customer dissatisfaction, and complexity of the source code.

In another study of the area, Ernst et al. \cite{ernst2015measure} performed an industrial survey to identify how software practitioners understand the TD concept and whether they use practices and tools for managing TD items. The study also considered a follow-up step with one-hour interviews for collecting data. In the survey, the authors asked that the participants order a predefined list of 14 causes of TD using as criterion the amount of debt that each cause could bring to the project. The survey received 536 responses, whereas the interviews were conducted with seven respondents. As a result, the authors identified that immature choices done in architecture are a key cause of TD.

Yli-Huumo et al. \cite{yli2015benefits} conducted a case study in two software organizations for identifying the benefits and consequences of taking workarounds. The data collection was performed by using semi-structured interviews with 17 practitioners. Open coding was used for analyzing the data and identifying categories related to the workarounds. The authors found that time pressure, complicated code base, no time for changing the selected software components, and prioritization of features based on their business value were the causes of workarounds. Also, the effects identified were decreased code maintainability, extra working hours, extra costs, major refactoring, lack of motivation to work with the code base, increased time for newcomers to start, outdated software components, lack of new features available in newer versions of components, and decreased code maintainability when scaling the feature.

Through a systematic mapping study, Li et al. \cite{li2015systematic} recognized that the following quality attributes are compromised due to the presence of debt: maintainability, reliability, security, portability, and performance efficiency. 

Concerning architectural TD, Martini et al. \cite{martiniEUROMICRO2014} performed a multiple case study including seven Scandinavian sites in five large software development organizations. The goal of the study was to investigate the causes for the accumulation of architectural TD and to investigate the trends in the accumulation and recovery of architectural TD. Through interactive workshops, the authors collected the data and analyzed it using open and axial coding, and a deductive approach. The authors identified that causes of TD are related to business factors, design and architecture documentation (lack of specification/emphasis on critical architectural requirements), reuse of legacy/third party/open source, parallel development, effects uncertainty, non-completed refactoring, technology evolution, and the human factor. In 2017, Martini and Bosh \cite{martiniATD2017} included two other Scandinavian sites and one large organization in that study to understand which items of architectural debt are the most dangerous for accumulated interest. The authors performed interviews and workshops, and identified a set of effects related to architectural TD, organizing those effects in a model. 

In another case study, Martini and Bosh \cite{martini2015towards} investigated the factors that should be considered during TD prioritization. They used questionnaires and interviews for collecting data. The authors reported that the effects of architectural TD are big deliveries, many code changes, “double” the effort, number/complement of tests, quality issues (bugs), wrong effort estimation, contagious architectural technical debt, probable hidden architectural technical debt, and developers idling. 

In summary, the current evidence on causes and effects of TD is limited to the point of view of a small set of practitioners not providing a complete view on causes that lead to the occurrence of TD and effects felt by practitioners due to the presence of TD. 

Technical debt can be injected into almost any artifact and during any stage of software development resulting with different types of technical debt---\textit{TD types} \cite{TD2018tertiaryStudy}---e.g., code, architecture, design, documentation, test, process, or people debt. Elaborate definitions of TD types are given in Section \ref{sec.Demographics.TDTypes} alongside with survey results.

\subsection{Previous Work}
\label{sec.PreviousWork}

This work is part of the InsighTD project, which intends to investigate the causes, effects, and management of TD \cite{OuluESEMrios2018most}.  Several papers have already been disseminated in the technical literature reporting the results of the project. In this subsection, we discuss our previous work on causes and effects of TD, and show how this work composes the big picture we are drawing with InsighTD.

Rios et al. \cite{OuluESEMrios2018most}  briefly presented the design of the project, along with the discussion of the top 10 causes and effects of TD recognized in the Brazilian software industry. After, the whole set of causes and effects from the Brazilian replication of the study was organized into probabilistic diagrams for supporting the analysis of TD causes and effects in software projects \cite{rios2019causesBrasil}. Based on this work, Rios et al. \cite{rios2020practitioners} reported a detailed discussion about the design of the InsighTD family of surveys and described the results of new analyses performed over the first execution of InsighTD in Brazil.

Replications of the survey have been performed in several countries and the initial analysis on causes and effects of TD have been incrementally reported. Perez et al. \cite{PerecBoris2020} identified the causes of TD in the Chilean software industry and Ramač et al. \cite{RamacEuromicro2020} recognized the causes and effects of TD in the Serbian software industry. Both studies compare their results to the ones reported by Rios et al. \cite{rios2019causesBrasil}. So far, considering only the results of the isolated replications’ results, we have investigated the following research questions: 
(1) \textit{Are software professionals familiar with the concept of TD?} 
(2) \textit{What causes lead software development teams to incur TD?}
and (3) \textit{What effects does TD have on software projects?}

However, a consolidated view of causes and effects of TD across the available\footnote{Currently, we already have the data from Brazil, Chile, Colombia, Costa Rica, Serbia, and the United States} replications of InsighTD is still missing. Moreover, until recently, the InsighTD dataset has not been large enough to make specific data analysis on causes and effects of TD. This paper represents the first opportunity to explore such specific analyses. 

For more information on the big puzzle we are working on within the InsighTD project, including other research variables, an interested reader can visit \url{http://www.td-survey.com/publication-map/}.

\section{Research Approach}
\label{sec.ResearchApproach}

The research approach 
encompasses the selection of the research strategy, organization, and execution of the methodological tasks. Consequently, the presentation of the research approach consists of: (a) selected research strategy and the argumentation in favor of the selection, (b) InsighTD project as a managing organization that guides the study, (c) design of the survey, which is the main research instrument, (d) survey family as a mode of survey execution, (e) data collection, with emphasis on sampling, (f) data analysis, presenting both quantitative and qualitative methods, and (g) the aggregation of the collected survey family data.   

\begin{table}[h]
    \caption{Argumentation for Survey Research Strategy in InsighTD}
    \label{tab.surveyResearchStrategyInsighTD}
    \centering
    \scriptsize
    \begin{tabular}{cp{5cm}p{5cm}}
        \toprule
        \textbf{No} & \textbf{Survey Research Strategy} & \textbf{InsighTD project} \\
        \midrule
        {1} & Survey can be used to ask question about what has happened in the past \cite{KitchenhamPfleegerLawrence,GhaziEtAllSurvey}. & Existing experience with TD, collected from IT practitioner, is in the focus of the analysis in InsighTD. \\
        \midrule
        {2} & Surveys can easily collect large number of variables \cite{Wohlin2006}. & Each RQ in InsighTD project is related with a subset of questions in the survey.   \\
        \midrule
        {3} & Survey allows to form an assertions (or claims) about target population \cite{babbie1973survey}   & Answer to RQ1 represent an assertion about the familiarity with the TD concept. \\
        \midrule
        {4} & Survey allows to form an explanation about target population \cite{babbie1973survey} & RQ2 and RQ3 represent explanations about the causes and effect of TD. \\
        \midrule
        {5} & Survey research allows the utilization of both qualitative and quantitative methods for analysis \cite{Wohlin2006}. & InsighTD relies on both qualitative and quantitative methods for analysis.  \\
        \midrule
        {6} &Surveys are cost effective, time effective, easy to distribute especially if the survey is implemented as web-based (or online) survey \cite{Lethbridge2005}. & In InsighTD, survey was implemented and distributed as online survey.  Cost and time effectiveness, and ease of distribution are useful properties. \\
        \bottomrule
    \end{tabular}
\end{table}

\subsection{Research Strategy}
\label{sec.ResearchStrategy}

The InsighTD project adopted the \textit{survey research strategy} \cite{KitchenhamPfleegerLawrence, MollieriEtAllSurveyGuidelines} as an overall methodological guidance for designing a family of surveys. In short, the survey research strategy consists of methodological steps that center on a survey or questionnaire\footnote{Both terms are used to denote the list of questions that participants are asked to answer} to collect the data from the sample of the population. Based on the analysis of that data, survey research aims to draw conclusions about the entire population \cite{babbie1973survey}, or to collect empirical data, which is a more attainable aim for survey research in the software engineering field. The decision to rely on this research strategy was made since it is seen as the most suitable for achieving the research objective. To argument this decision, Table \ref{tab.surveyResearchStrategyInsighTD} presents how some of the main properties of the survey strategy are related with InsighTD.

The implementation of the survey research strategy in InsighTD followed the recommendations for survey research in software engineering proposed by Mollieri et all. \cite{MollieriEtAllSurveyGuidelines}. The following sections present the adoption of these recommendations in InsighTD. To facilitate the presentation, these sections refer to Figure \ref{fig.InsTDorganization} which relates the organization of InsighTD with the main research activities.

\subsection{The InsighTD Project and Research Questions}
\label{sec.InsTdProject}

\begin{figure}[h]
    \centering
 \includegraphics[width=12cm, height=10.34cm]{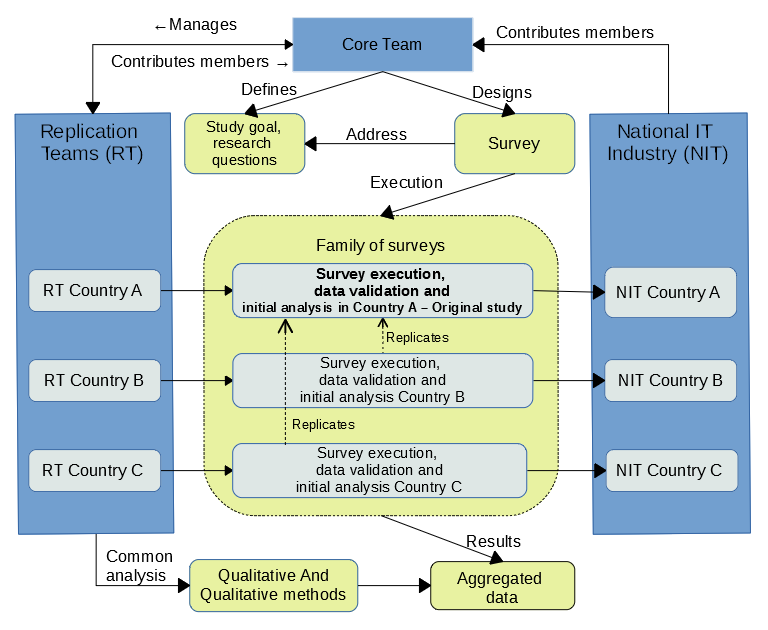}
    \caption{InsighTD project organization}
    \label{fig.InsTDorganization}
\end{figure}

The InsighTD project is an international initiative that focuses on the TD phenomenon and its manifestation in the software development process. The project organization consists of the core team and national replication teams. The core team is an international group of senior scientists who led the project preparation and made all relevant decisions regarding the project organization and research. The core team also coordinates the efforts of national replication teams. Replication teams are groups of scientists from each country that participates in InsighTD. At the time of writing this article, 34 researchers organized in national teams from 12 countries joined the initiative\footnote{Project statistics on 28.Nov.2020}. The main responsibility of the replication teams is to execute the InsighTD tasks in their own countries but also to facilitate the work of the core team, or to contribute members to the core team. In Figure \ref{fig.InsTDorganization}, the core and replication teams are presented with rectangles, whereas the replication teams are additionally emphasized as rounded rectangles.  

The core team has set the foundation of the InsighTD project by defining the goal, research questions and by selecting the overall research strategy. The goal of InsighTD is to produce generalizable knowledge on the TD phenomenon and its manifestation in the software development process. To achieve this goal, the project focuses on several aspects of TD, three of these aspects were in the focus of this study and articulated in the following research questions: 

\begin{itemize}
\item[\textbf{RQ1:}] To what extent are software professionals familiar with the concept of TD?
\item[\textbf{RQ2:}] What causes lead software development teams to incur TD? 
\item[\textbf{RQ3:}] What effects does TD have on software projects? 
\end{itemize}

InsightTD seeks to investigate the causes of TD, without differentiating whether the debt is incurred intentionally or unintentionally, and the effects without differentiating whether they positively or negatively impact the development process. For this reason, RQ2 and RQ3 remained agnostic with respect to these aspects, so were the survey participants since they were not instructed to consider them either.  

Number of InsighTD studies that focus on one or more of the presented RQs have already been published within the InsighTD project\footnote{http://www.td-survey.com/publication-map/}. These studies, however, relied on partial datasets, e.g. data collected in a single country, or data collected in companies using agile approach. This study, is the first study whose results are based on an aggregated dataset, and thus brings us closer to generalizable knowledge on TD which is the goal of the InsighTD project.

\subsection{Motivation for, and Design of InsighTD Family of Surveys}
\label{sec.SurveyDesignFamily}

Software engineering recognizes two major approaches for analyzing aggregations of empirical studies \cite{shull2008role,santos2019procedure}: (1) systematic literature reviews (SLR), i.e. meta-studies of published research, and (2) replications of empirical studies, i.e. a \textit{family of empirical studies}. The advantages of replications over SLRs are: 
\begin{enumerate}
\item Replications share the same research goal, and if they share the same research questions then we call them exact replications \cite{santos2018analyzing,santos2019procedure}.  
\item Replications are conducted by independent researchers or research teams.
\item Research design (research protocol and instruments) is shared by all replications, which allows for the collection of data in a uniform and consistent way.
\item Often, data analysis protocols are shared by replications.
\item All researchers have access to raw data, unlike SLRs, and therefore an aggregated data analysis is possible.
\item Joint conclusions are not affected by publication bias \cite{santos2019procedure}.
\end{enumerate}

The replications, that is \textit{families of empirical studies}, are therefore perceived to have higher internal validity over SLRs. Families of surveys in the SE field are still rare and one example is from the requirements engineering domain \cite{wagner2019status}. To the best of our knowledge, InsighTD is the first family of surveys in the TD domain.

In survey research, the questionnaire (or the survey) is a central research instrument and it is used to collect data from the target population \cite{babbie1973survey}. The survey used in InsighTD was designed by the core team, and the details about this design are already presented in scientific publications \cite{OuluESEMrios2018most, rios2020practitioners}. The following sections present the most relevant information regarding the survey and its design.

The InsighTD survey\footnote{Survey description and the final version of the survey is available on following link: \url{https://bit.ly/2SCOM9S}} consists of 28 questions and these questions are organized into five parts. The first part are the demographics questions that are used to characterize the respondents, while the subsequent three parts aim to elicit the data for answering the three stated RQs respectively. Fifth part targets TD management aspects that are not part of this study. Demographic questions are all closed-type questions, while the other parts use both closed and open-ended type questions. A subset of survey questions related to RQs addressed in this study are presented in Table \ref{tab.Questionnaire}. To design a survey, the core team relied on the existing knowledge from scientific sources and on insights from senior researchers and industry experts, as presented in \cite{OuluESEMrios2018most}.  

\begin{table}[H]
\centering
  \caption{Subset of the InsighTD Survey’s Questions for demographics, RQ1, RQ2, and RQ3. Adapted from \cite{OuluESEMrios2018most}}
  \label{tab.Questionnaire}
 \scriptsize
  \begin{tabular} {ll p{7cm} l}
 
    \toprule
    \textbf{RQ} & \textbf{Qn} & \textbf{Question} & \textbf{Type} \\
    \midrule
Demographics & Q1 & What is the size of your company? & Closed \\
Demographics & Q2 & In which country you are currently working? & Closed \\
Demographics & Q3 & What is the size of the system being developed in that project? (LOC) & Closed \\
Demographics & Q4 & What is the total number of people of this project? & Closed \\
Demographics & Q5 & What is the age of this system up to now or to when your involvement ended? & Closed \\
Demographics & Q6 & To which project role are you assigned in this project? & Closed \\
Demographics & Q7 & How do you rate your experience in this role? & Closed \\
Demographics & Q8 & Which of the following most closely describes the development process model you follow on this project? & Closed \\
RQ1 & Q9 & How familiar you are with the concept of Technical Debt? & Closed \\
RQ1 & Q10 & In your words, how would you define Technical Debt? & Open \\
RQ1 & Q11 & How close to the above TD definition is your understanding about TD? & Closed \\
RQ1 & Q12 & Are there any parts of the definition above from McConnell that you disagree with? Are there some things that you think are TD that are not included in this definition? & Open \\
RQ1 & Q13 & Please give an example of TD that had a significant impact on the project that you have chosen to tell us about: & Open \\
RQ1 & Q14 & Why did you select this example? & Open \\
RQ1 & Q15 & About this example, how representative it is? & Closed \\
RQ2 & Q16 & What was the immediate, or precipitating, cause of the example of TD you just described? & Open \\
RQ2 & Q17 & What other cause or factor contributed to the immediate cause you described above? & Open \\
RQ2 & Q18 & What other motives or reasons or causes  contributed either directly or indirectly to the occurrence of the TD example? & Open \\
RQ2 & Q19 & Considering all the cases of TD you've encountered in different projects, and the causes of those TD cases, which causes would you say are the most likely to lead to TD (ordered by likelihood of causing TD)? Please list up to 5 causes. & Open \\
RQ3 & Q20 & Considering the TD item you described in question 13, what were the impacts felt in the project? & Open \\
RQ3 & Q21 & Considering all the cases of TD you've encountered in different projects and the effects of that TD that you have personally experienced, which 5 effects would you classify as the effects that have a bigger impact (ordered by their level of impact). & Open \\
    \bottomrule
  \end{tabular}
\end{table}

Once the survey was designed, replication teams were free to execute the survey as is, or they could introduce minor adaptations. For example, if needed, replication teams translated the questions to the national language and allowed answers to open-ended questions to be both in English and in the national language. The reasons for these adaptations were the insights from the pilot study which suggested that they can improve the quality of the responses \cite{mandicTD2020}. This ensured that the language barrier was neutralized and that the survey can now be conducted in any country that joins the InsighTD initiative and that the aggregated analysis could now be complemented with data from various countries.

Following the methodological steps that are set in the InsighTD project, the designed survey was executed as a \emph{family of surveys}. In InsighTD, a family of surveys was realized by independent execution of surveys and analysis by replication teams. Each replication team executed the survey in their country, that is in their national IT industry and followed the identical procedure for the analysis. In the central part of Figure \ref{fig.InsTDorganization}, the survey family is presented with a dotted-lined rounded rectangle which contains blocks for each survey executions of each replication team within their national IT industry. Further more, Figure \ref{fig.InsTDorganization} also shows that replicated surveys addressed the same study goal and RQs, and that the data is aggregated in a common data repository and then analyzed.  

The first survey in this survey family was executed by the Brazilian team \cite{OuluESEMrios2018most}, thus this study is referred as the \textit{original study}. All subsequent survey executions, executed by other replication teams, replicate the original study and therefore are referred as the replication studies. In Figure \ref{fig.InsTDorganization}, the block with the original study is emphasized with bold letters. The replication studies were presented below and related with the original study with the dotted arrows. 
The survey was implemented as an online survey.

\subsection{Data Collection}
\label{sec.DataCollection}

TD represents a concept specific to software engineering practice, hence the target population for this survey consists of IT practitioners working in the software development industry. The sampling from this target population was done following the \textit{convenience sampling} approach \cite{wohlin2012experimentation, Baltes2020}. In short, convenience sampling is a non-probability sampling approach where the sample is drawn based on convenience, or availability of the subjects. This sampling approach was seen as suitable since any IT practitioner in any role working in the IT industry was seen as a suitable candidate for the survey. Consequently, each replication team sent survey participation invitations using all or some of the following channels: (a) invitation using their personal contacts who work in the IT industry, (b) asking their contacts to further disseminate the invitation to their colleagues, (c) in some countries, IT cluster organizations were contacted and asked to distribute the invitation among their members, and lastly (d) IT practitioners were invited by posting the invitation to specialized forums using internet-based platforms and, such as LinkedIn\footnote{www.linkedin.com} or Facebook\footnote{www.facebook.com}.

Upon survey execution, each replication team was responsible for validating the data. The validation of open-ended questions was done by checking (i) the completeness of the answers and (ii) did the respondents provide a valid TD example. The completeness check was done by making sure that the answers properly address the questions. The evaluation of the TD example focused on the answer to Q13 where the respondents were asked to describe an example of TD. Depending on whether the provided example could be related with one of the TD types, the answers were used or discarded during latter analysis. This evaluation revealed whether the respondents properly understand the TD phenomenon and therefore increased the credibility of their answers. Each answer was read by at least three scientists, and to discard the answer that was seen as invalid, a consensus of at least two scientists was required. Finally, the validation of closed questions was ensured by constraints set by the online survey, and it was publicly accessible since mid 2018 until the end of 2019. As the online survey tool \textit{Google forms} were used.

\subsection{Data Analysis}
\label{sec.DataAnalysis}

The data is analyzed using both quantitative and qualitative research methods. Descriptive and median statistics, alongside with different charting and diagramming techniques, were the most frequently used quantitative methods in replication studies.
In this study, descriptive statistics with diagrams and charts were used to analyze the closed-type questions. The respondents answered these questions by selecting one of the choices that were offered in form of categorical, ordinal or interval scale and these choices were counted and analyzed. Additionally, the Pearson $\chi^2$ test with contingency tables was used to analyze the relations between the pairs of observed variables. Relations that were seen as interesting, i.e. statistically significant, were presented with mosaic plots. Mosaic plots with shading \cite{brown1992friendly} were chosen as a representation tool due to their expressiveness which allows them to show both the relation between two categorical variables, and the strength of that relation. To calculate the $\chi^2$ statistics and generate the mosaic plots, the R\footnote{https://www.r-project.org/} statistical tool was used.   

The answers to open-ended questions were analyzed using the \textit{qualitative coding technique}, which is one of the qualitative methods \cite{miles1994}. In short, coding denotes the process of labeling pieces of text which are seen as interesting with a meaningful name, i.e. the \textit{code}. This analysis consisted of three main steps. The first step was the thorough familiarization with the survey answers. The second was the \textit{coding} of the answers. This step is also refereed to as open coding, since everything that is seen as interesting was coded. The third step is the grouping of codes to \textit{categories}. During this step, the open codes are re-read and related based on their similarities to a higher-order code, that is to its category. This step is also refereed as axial coding. Each national team coded the dataset collected in their national context. To derive the codes and categories, a consensus of at least two scientists was required\footnote{Analysis procedure that is used for RQ1 is available online: \url{https://bit.ly/3fpfYS7}. Readers can familiarize with the level of details, and explanations in analysis procedures.}. Resulting codes, categories and their relations were described and interpreted.

The execution of the third step however differed in the original study and in replications. The third step in the original study consisted of axial coding, while in replications, besides axial coding, replication teams had to map their codes to the ones derived in the original study. To map the codes the replication teams relied on code descriptions from the original study and on the consensus among replication team members to confirm the mapping. This way, the list of codes and categories evolved with each replication by confirming existing and by adding new codes and categories to the list from the original study. 
The consolidated lists of codes and categories for TD causes and effects are given in Sections \ref{sec.Rq2.conceptualizatitonTDcauses} and \ref{sec.RQ3.conceputalizationEffects}. 

\subsection{Aggregation of Data and Results}
\label{sec.DataAggregation}
The data collected in different countries is aggregated into a single common dataset, however the aggregation differed for answers on closed and open questions. The answers to closed questions were merged into a common dataset as is, since these questions offer predefined answers to choose from. Only in cases when the survey is translated to the national language, the chosen answers were first translated to English, and then merged into a common set. The answers to open-ended questions however, are in free form text. This text was provided in national languages or in English. Therefore, instead of merging the raw data, each replication team completed the analysis of the data, and then contributed the analysis results into a common, consolidated dataset. This was possible, since each replication relied on the initial coding scheme that was derived in the original study, thus the results of the replication studies confirmed the existing codes and possibly further enriched them with new ones. 

For example, to aggregate the answers to open-ended questions relevant for answering RQs stated in this study, replications' data from Brazil, Chile, Columbia, Costa Rica, United States and Serbia where used. Each replication team analyzed the relevant answers, identified, and described codes and categories. The provided descriptions made the meaning of codes and categories understandable to each replication team, which in turn facilitated the merger of codes and categories into a consolidated dataset. This consolidated set of data was used to produce the results that are presented in the following section. 

\section{Results}
\label{sec.Results}
The data collection resulted with 745 responses from practitioners that came from Brazil, Chile, Colombia, Costa Rica, Serbia and the United States. Table \ref{tab.ParticipantsPerCountry} shows the number of participants per country. The column \textit{responses} represents the number of practitioners that answered the survey in the respected country, while the column \textit{valid responses} represents the number of participants that remained after the data was validated. In total, 12.35\% of the dataset was discarded, leaving 653 valid responses.

\begin{table}[h]
    \caption{Participants per country}
    \label{tab.ParticipantsPerCountry}
    \centering
    \scriptsize 
    \begin{tabular}{llll}
        \toprule
        Replication & Responses & Valid Resp. & Valid Resp. (\%)\\
        \midrule
        {Brazil}        & 112    & 107   & 16.39\% \\
        {Chile}         & 104    & 89    & 13.63\% \\
        {Colombia}      & 159    & 134   & 20.52\% \\
        {Costa Rica}    & 156    & 145   & 22.21\% \\
        {Serbia}        & 91     & 79    & 12.10\% \\
        {United States} & 123    & 99    & 15.16\% \\
        \midrule
        \textbf{Total}  & 745    & \textbf{653} & 100\%\\
        \bottomrule
    \end{tabular}
\end{table}

The study results are based on an aggregated analysis that took into account valid responses from participants from each country. The countries had the following share of the total number of valid responses: Brazil 16.39\%, Chile 13.63\%, Colombia 20.52\%, Costa Rica 22.21\%, Serbia 12.10\% and the United States 15.16\% (Table \ref{tab.ParticipantsPerCountry}).

The demographics section presents the exemplar project in Section \ref{sec.Demographics.ExemplarProject}, participants in Section \ref{sec.Demographics.Participants} and types of TD in Section \ref{sec.Demographics.TDTypes}. 
Section \ref{sec.Rq1.TDFamiliarity} reveals the identified relations between various demographic variables and the familiarity with the TD concept. Finally, the identified causes and effects were related with the TD types and this relation is presented in Sections \ref{sec.Results.TDCausesEffects}.
The presentation of results also includes the authors' interpretation of the results in the form of enumerated findings.

\subsection{Demographics data}
\label{sec.Demographics}

To collect the demographic data, survey questions Q1--Q8 were used, and these questions where all of closed type. Before the start of the survey, however, participants were also instructed to think of an exemplar project. The exemplar project would then serve as a context or as a reference point to which other survey questions would refer to.   

\subsubsection{Exemplar project}
\label{sec.Demographics.ExemplarProject}
The exemplar project is a project in which the participant was involved and which he or she deems as representative from the TD point of view. Participants were instructed that having a clear idea about the exemplar project is an important part of the survey, since the majority of answers had to be provided from the perspective of this project\footnote{Survey description and the final version of the survey is available on following link: \url{https://bit.ly/2SCOM9S}}. For example, in Q13, participants were asked to describe the example of TD that they encountered in the exemplar project. Later during the analysis, the TD example was mapped to one of the TD types. Depending on whether the TD example could, or could not be mapped to a TD type, the answer was accepted or discarded from further analysis. 

\begin{figure}
  \centering
  \includegraphics[width=10cm, frame]{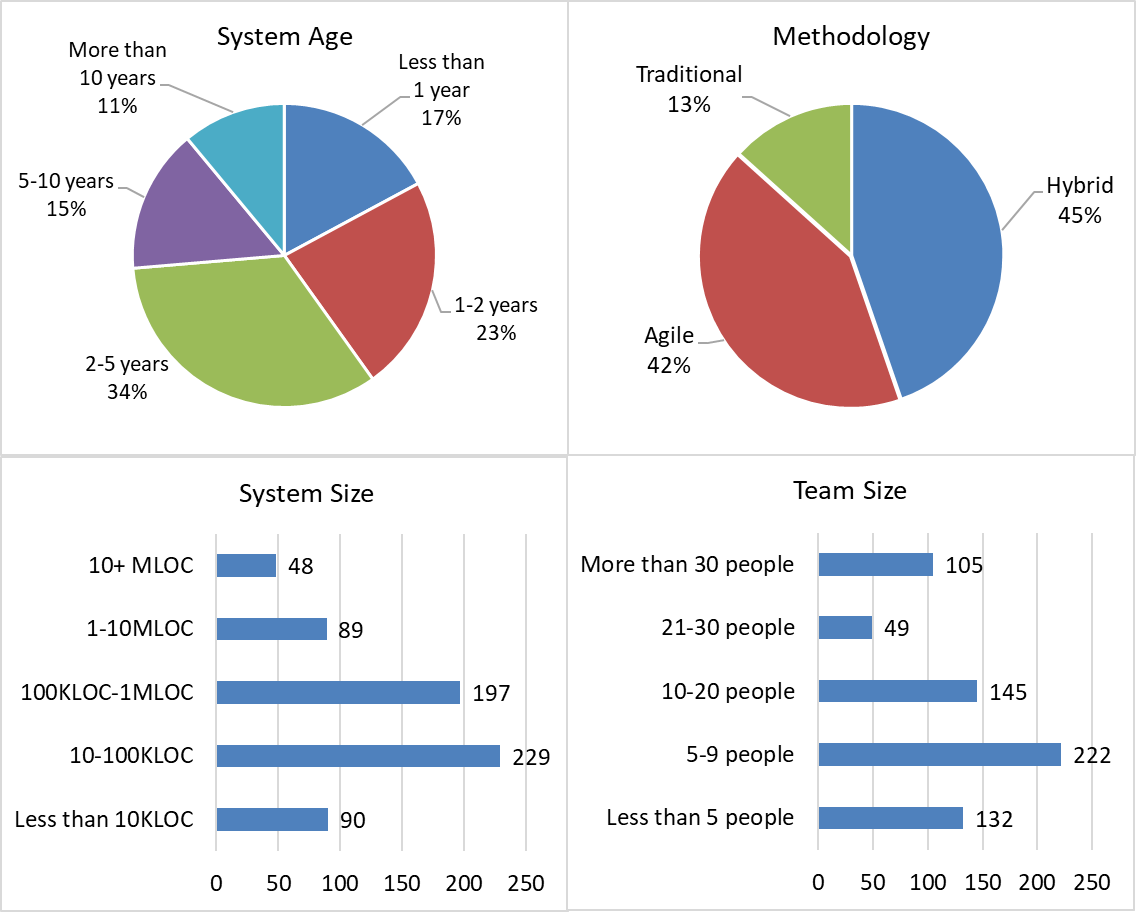}
  \caption{Demographics data for the exemplar project ($N=653$).}
   \label{fig.Demographics}
\end{figure}

The exemplar project is characterized with data regarding the development approach, system age, system and team size. This data is illustrated in Figure \ref{fig.Demographics}. The most dominant development approaches used were the hybrid approach (44.72\%) that consisted of combining agile approaches with more traditional development methodologies (e.g. waterfall), and the agile approach (41.96\%). The remaining 13.32\% of the participants stated that some traditional methodology was used in the development of the exemplar project.

More than half of the specified systems were between 1 and 5 years of age (56.51\%), while others were scattered across less that 1 year old (17.15\%), between 5 and 10 years old (15.31\%) and older than 10 years of age (11.03\%).

Regarding system size, most of the systems were between 10KLOC and 1MLOC (65.24\%). Systems with less than 10KLOC took up 13.78\% of all systems while systems between 1MLOC and 10MLOC took up 13.63\%. Systems larger than 10MLOC accounted for only 7.35\% of all specified systems in the survey.

Teams that worked on the exemplar project ranged from teams that counted less than 5 people to large teams that had more than 30 people. The majority of teams counted between 1 and 20 people (76.42\%) while the remaining 23.58\% was distributed across teams that counted from 21 to 30 people (7.5\%) and more than 30 people (16.08\%).

\subsubsection{Participants} 
\label{sec.Demographics.Participants}

During the survey, the participants were asked to chose the role they had in the exemplar project from the provided list, or to specify the role if it was not offered in the list. The provided list of roles included for example: developer, tester and software architect roles. The complete list of proposed roles is enumerated in Table \ref{tab.RoleCategories}. Afterwards, the participants were asked to specify the level of experience they had while working in the specified role. The experience scale included: Novice (minimal or textbook knowledge without connecting it to practice), Beginner (working knowledge of key aspects of practice), Competent (good working and background knowledge of area of practice), Proficient (depth of understanding of discipline and area of practice), and Expert (authoritative knowledge of discipline and deep tacit understanding across area of practice). 

The results show that in total there were (note that the percentages were rounded): 2\% novice, 12\% beginner, 34\% competent, 32\% proficient, and 21\% expert participants. It can be concluded that only 14\% of the participants can be classified as juniors (novice and beginner) that have little experience, while 87\% can be classified as experienced in their roles. Table \ref{tab.Demographics} shows the distribution of different experience levels over various roles. The roles marked with an asterisk (*) are roles that were not predefined in the InsighTD questionnaire, thus represent the roles defined by participants. These role were bundled into the \textit{other} category. The most dominant role is the developer\footnote{the developer role often spans over the entire software development process, e.g. it often includes system design and architecture activities, testing and so on.} role that accounts for 50\% of all respondents. The majority of respondents for this role can be classified as experienced (84.71\%).

\begin{table}[h]
    \caption{Participants' roles and experience levels per role. R1--Developer; R2--Project leader/manager; R3--Software architect; R4--Tester; R5--Requirements analyst; R6--Process analyst; R7--Database administrator/analyst; R8--Business analyst; R9*--Performs multiple functions; R10*--Quality analyst; R11*--Configuration manager; R12*--Infrastructure analyst; R13*--Infrastructure manager; R14*--Quality analyst manager}
    \label{tab.Demographics}
    \centering \scriptsize
        \begin{tabular}{c
                        *{2}{S[table-format=0.5]}
                             S[table-format=2.0] 
                        *{3}{S[table-format=1.0]}
                             S[table-format=2.0]
                       }
        \toprule
        \multicolumn{2}{c}{} &
        \multicolumn{2}{c}{Junior (\%)}
            & \multicolumn{1}{c}{}    
            & \multicolumn{3}{c}{\makecell{Experienced (\%)}}\\
        \cmidrule{3-4}
        \cmidrule{6-8}
        {Role} & {N (\%)} & {Novice} & {Beginner} & & {Competent} &  {Proficient} & {Expert}  \\
        \midrule
        \makecell[l]{R1}   &{327 (50.08)} &1.83  &13.46  &   &37.00  &30.28  &17.43  \\
        \makecell[l]{R2}   &{108 (16.54)} &0.93  &8.33   &   &29.63  &35.19  &25.93  \\
        \makecell[l]{R3}   &{87 (13.32)}  &1.15  &8.05   &   &24.14  &40.23  &26.44  \\
        \makecell[l]{R4}   &{46 (7.04)}   &0.00  &6.52   &   &34.78  &34.78  &23.91  \\
        \makecell[l]{R5}   &{24 (3.68)}   &0.00  &20.83  &   &45.83  &25.00  &8.33   \\
        \makecell[l]{R6}   &{19 (2.91)}   &0.00  &31.58  &   &36.84  &15.79  &15.79  \\
        \makecell[l]{R7}   &{14 (2.14)}   &0.00  &21.43  &   &28.57  &28.57  &21.43  \\
        \makecell[l]{R8}   &{13 (1.99)}   &7.69  &15.38  &   &30.77  &15.38  &30.77  \\
        \makecell[l]{R9*}  &{5 (0.77)}    &0.00  &20.00  &   &40.00  &20.00  &20.00  \\
        \makecell[l]{R10*} &{3 (0.46)}    &0.00  &0.00   &   &33.33  &33.33  &33.33  \\
        \makecell[l]{R11*} &{2 (0.31)}    &0.00  &0.00   &   &0.00   &0.00   &100.00 \\
        \makecell[l]{R12*} &{2 (0.31)}    &50.00 &0.00   &   &50.00  &0.00   &0.00   \\
        \makecell[l]{R13*} &{2 (0.31)}    &0.00  &0.00   &   &0.00   &50.00  &50.00  \\
        \makecell[l]{R14*} &{1 (0.15)}    &0.00  &0.00   &   &0.00   &0.00   &100.00 \\
        \bottomrule
        \end{tabular}
\end{table}

Regarding company size, the participants are somewhat equally distributed into the following company sizes: 32.16\% in small companies that count up to 50 employees, 38.44\% in medium sized companies that have 51 to 1000 employees and the remaining 29.4\% work in large companies that count more than 1000 employees.

\subsubsection{Technical Debt Types}
\label{sec.Demographics.TDTypes}

\begin{figure}
  \centering
  \includegraphics[width=8cm, frame]{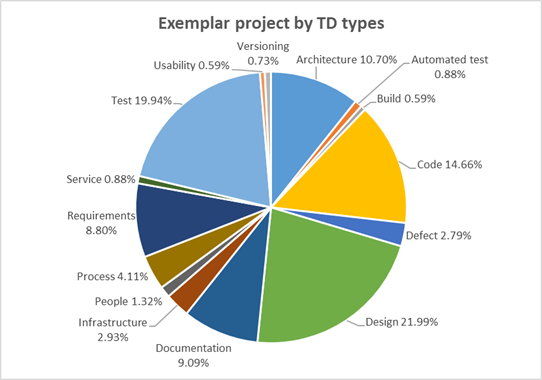}
  \caption{Exemplar TD Types.}
   \label{fig.ExemplarTDTypes}
\end{figure}

Types of TD were derived based on the exemplar project during which the survey participants recognized or experienced situations that led to the injection of TD. The analysis of valid responses resulted with the identification and extraction of the TD Types from the described situations.   
Figure \ref{fig.ExemplarTDTypes} depict the distribution of TD Types.

\textit{Design debt} ($21.99\%$) refers to violations of good design principles and practices. This type of debt can be discovered by analyzing the source code. For example, using code review techniques or static code analysis tools \cite{li2015systematic}.

\textit{Test debt} ($19.94\%$) refers to any issues in regard with testing activities \cite{guo2011portfolio}. For example, planned testing that was not conducted or low-test coverage.

\textit{Code debt} ($14.66\%$) refers to problems found in the source code that negatively affects readability and maintainability of the code. For example, too complex methods or non-existing coding standards. In order to remove code debt, code refactoring is needed \cite{TD2018tertiaryStudy}.

\textit{Architecture debt} ($10.70\%$) refers to issues and problems with the architecture of the system that affect the architectural requirements (e.g., performance, maintainability, robustness, etc.) \cite{kruchten2012technical}. For example, violation of the modularity principle or poor separation of concerns leads to significant problems with further evolution, maintenance and extensions of the system.

\textit{Documentation debt} ($9.09\%$) refers to any issues and problems discovered in the documentation of the software project, including missing documentation. It is discovered by looking for inconsistent, missing, inadequate or incomplete documentation \cite{guo2011portfolio}. For example, technical documentation is out dated and not reflecting the actual state of the system.

\textit{Requirements debt} ($8.80\%$) refers to the trade off between agreed requirements and requirements that are actually implemented in a software product. For example, this can be manifested as partially implemented requirements or requirements that are implemented for some cases only \cite{kruchten2012technical}. 

\textit{Process debt} ($4.11\%$) refers to inefficient processes. For example, a process changed over time and the existing practices are not efficient any longer \cite{TD2016identification}.  

\textit{Infrastructure debt} ($2.93\%$) refers to issues that if present in an organization significantly reduce the teams’ abilities to deliver quality products \cite{TD2016identification}. For example, using outdated tools or postponing regular software or system updates.

\textit{Defect debt} ($2.79\%$) refers to known defects that are discovered by testing activities, however the corrective actions are postponed or never implemented. Decisions to postpone the corrective actions can accumulate a significant amount of TD in the system \cite{snipes2012defining}.

\textit{People debt} ($1.32\%$) refers to any issues related with people. For example, lack of specific skills and knowledge that would require additional training or hiring \cite{TD2016identification}. People related issues are much more complex, they involve socio-organizational factors that directly affect productivity and people satisfaction \cite{TD2018tertiaryStudy,burnoutIEEESw2019}.

\begin{finding}
\label{fin.TDTypes}
There is no single TD type that stands out. The top-5 debt types by its occurrence are: design,  test,  code, architecture, and documentation.
\end{finding}

Other debt types had less than 1\% occurrences each: \textit{service} ($0.88\%$), \textit{automated test} ($0.88\%$), \textit{versioning} ($0.73\%$), \textit{usability} ($0.59\%$), and \textit{build} ($0.59\%$). Precise definitions for these debt types can be found in \cite{mandicTD2020}. 

\subsection{The familiarity with the TD concept (RQ1)}
\label{sec.Rq1.TDFamiliarity}
\begin{figure}
  \centering
  \includegraphics[width=12cm, height=4cm, frame]{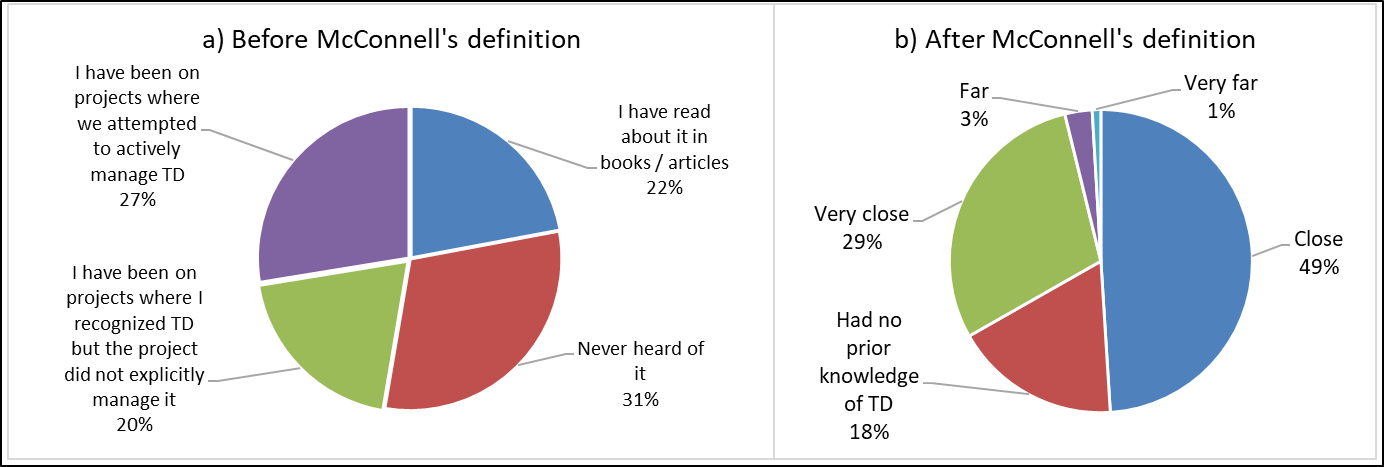}
  \caption{Participants' familiarity with the TD concept: a) before TD definition, b) after TD definition ($N=653$).}
   \label{fig.RQ1}
\end{figure}

To assess the extent of familiarity with the TD concept, participants were asked to self-assess their own familiarity with TD. Afterwards they were given the McConnell's definition and asked to specify how close is their understanding of TD to the definition. 
Figure \ref{fig.RQ1}.a shows that 69\% of participants were familiar with the TD concept and that 31\% of the participants have never heard of it. Out of 69\% of the participants that were familiar with TD, 22\% had only a theoretical knowledge (they read about the TD concept in books or articles), whereas 47\% had more practical experience and were either on projects where they actively managed TD or TD was recognized but not managed.

Figure \ref{fig.RQ1}.b shows that a large percentage of the participants specified that the McConnell's definition was either close or very close to their understanding of TD (78\%), while 4\% of the participants seem to disagree with the definition. From Figure \ref{fig.RQ1}.b it can also be seen that the percentage of participants that had no prior knowledge of TD decreased to 18\%, compared to 31\% of respondents that answered that they have never heard of TD before the definition was revealed. This indicates that after presenting the McConnell's definition, the TD concept resonated with some of participant's past experiences that prior to the definition revelation answered that they have never heard of TD before.  

\begin{finding}
\label{fin.Finding1Prevalence}
The large majority of practitioners understands or agrees with the McConnell's definition of TD (78\%).  Whereas, only 47\% had some practical experiences with TD identification or management. Furthermore, the definition resonated well with practitioners’ experiences who had no prior knowledge about the TD concept. 
\end{finding}

In order to gain better insights on how demographics and TD familiarity interact additional analyses were conducted. For every demographics variable a two step analysis was done. First, a $\chi^2$ test was conducted to see if there is a statistically significant association (dependency) between demographic variables on one side, and TD familiarity and the apprehension of McConnell's definition on the other. This was possible because the variable under study is categorical and the dataset is relatively large. To conduct this test the R language was used with the help of the \textit{stats} package\footnote{https://www.rdocumentation.org/packages/stats/versions/3.6.2/topics/chisq.test}, all in combination with the R Studio development environment. Second, if the test showed 
the existence of statistically significant associations then mosaic plots with shading were used to visualize data associations.

\begin{table}[H]
    \caption{Results of a $\chi^2$ test between demographics data and TD familiarity (Q9) and TD definition compliance (Q11). $\chi^2$--chi square test results; df--degrees of freedom; p-value--threshold $\alpha=0.05$}
    \label{tab.DemographicsTDFamiliarity}
    \centering \scriptsize
    \begin{tabular}{llllllll} 
        \toprule
        & \multicolumn{3}{c}{Q9: Familiarity}                                                          
        &  & \multicolumn{3}{c}{Q11: Apprehension}\\ 
        \cline{2-4}\cline{6-8}
        \begin{tabular}[c]{@{}l@{}}Demographic\\variable\end{tabular} & 
            \multicolumn{1}{c}{$\chi^2$} & 
            \multicolumn{1}{c}{df} & 
            \multicolumn{1}{c}{p-value} &  & 
            \multicolumn{1}{c}{$\chi^2$} & 
            \multicolumn{1}{c}{df} & 
            \multicolumn{1}{c}{p-value}  \\ 
            \midrule
                {Country}       &87.912  &15  &{2.428e-12}  &   &35.658  &20  &0.01687     \\
                {Company size}  &13.865  &6   &0.03118  	&   &6.6222  &8   & \textcolor{red}{0.5779}      \\
                {System size}   &37.586  &12  &0.0001794  	&   &28.379  &16  &0.02848     \\
                {Team size}     &47.547  &12  &{3.746e-06}  &   &20.757  &16  &\textcolor{red}{0.1881}      \\
                {System age}    &15.192  &12  &\textcolor{red}{0.2311}  &   &17.802  &16  &\textcolor{red}{0.3356}      \\
                {Role}          &96.746  &39  &{8.229e-07}  &   &85.1  	 &52  &0.002567    \\
                {Experience}    &70.369  &12  &{2.732e-10}  &   &71.783  &16  &{4.851e-09} \\
                {Methodology}   &39.688  &6   &{5.246e-07}  &   &11.791  &8   &\textcolor{red}{0.1608}      \\
            \bottomrule
    \end{tabular}
\end{table}

Table \ref{tab.DemographicsTDFamiliarity} shows that no significant association ($p-value>0.05$, red font)  was found between system age and TD familiarity, as well as between company size, team size, system age and methodology and the apprehension of McConnell's definition. The following text present the in depth analysis of the significant associations (p-values in black font) between the demographics variables and the familiarity with the TD concept. The associations between the demographic variables and the apprehension of the McConnell's definition were not analyzed regardless of the statistical significance. In depth analysis result of the apprehension of the TD definition may contribute to areas such as the IT education or similar, however it is the authors' opinion that this analysis offers no contribution to answers for RQ1 and it was therefore omitted from the study.

\begin{figure}
  \centering
  \includegraphics[width=12cm, height=6.5cm, frame]{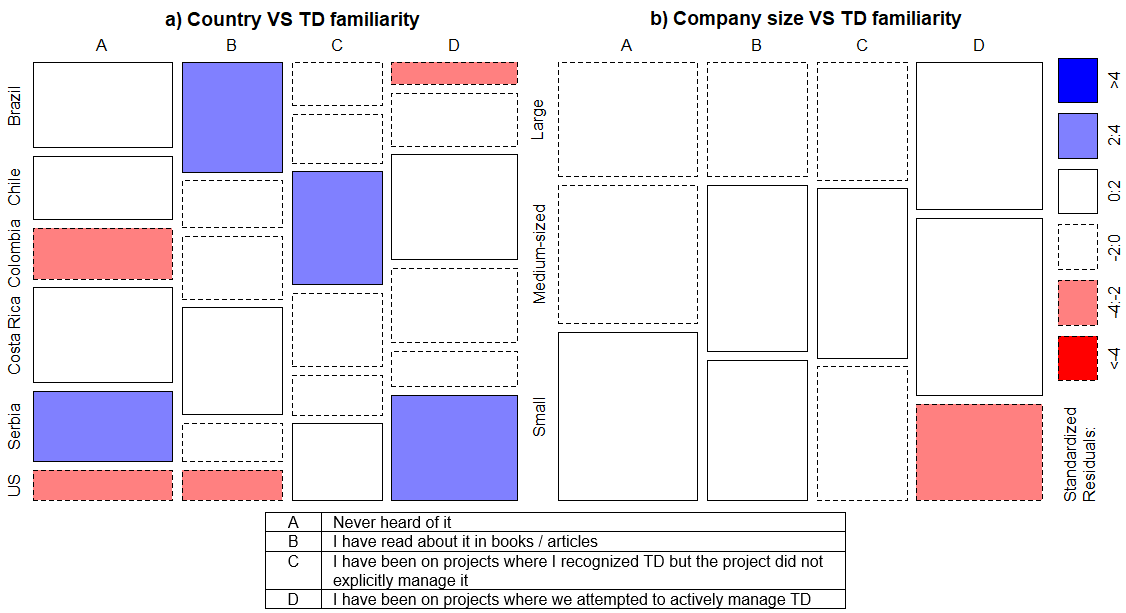}
  \caption{The mosaic plot for country and company size versus TD familiarity (Q9)}
   \label{fig.mosaic.CountryCompanySizeVSQ9}
\end{figure}

\paragraph{Country versus TD familiarity}
Figure \ref{fig.mosaic.CountryCompanySizeVSQ9}.a depicts a mosaic plot of prevalence of the TD concept in industry, with respect to country. Analysis of the figure shows that practitioners that come from Serbia are more likely to have never heard of the TD concept. If practitioners come from the United States or Colombia it is of low probability that they never heard of the TD concept, and regarding United States practitioners they are more likely to have had some practical experience in managing TD. On the other hand, practitioners from Colombia are more likely to have practical experience in TD identification. Brazilian practitioners are more likely to have theoretical knowledge of the TD concept and less likely to have had some practical experience in managing TD.

\begin{finding}
\label{fin.Finding2CountryVSPrevalence}
Practitioners coming from different countries may have different levels of familiarity with the TD concept. For example, practitioners from USA or Colombia are more likely to be familiar with TD, whereas practitioners from Serbia are not.
\end{finding}

\paragraph{Company size versus TD familiarity}
Figure \ref{fig.mosaic.CountryCompanySizeVSQ9}.b depicts the prevalence of the TD concept in industry, with respect to company size and shows that practitioners who come from small companies are less likely to have practical experience with TD management.

\begin{figure}
  \centering
  \includegraphics[width=12cm, height=6.5cm, frame]{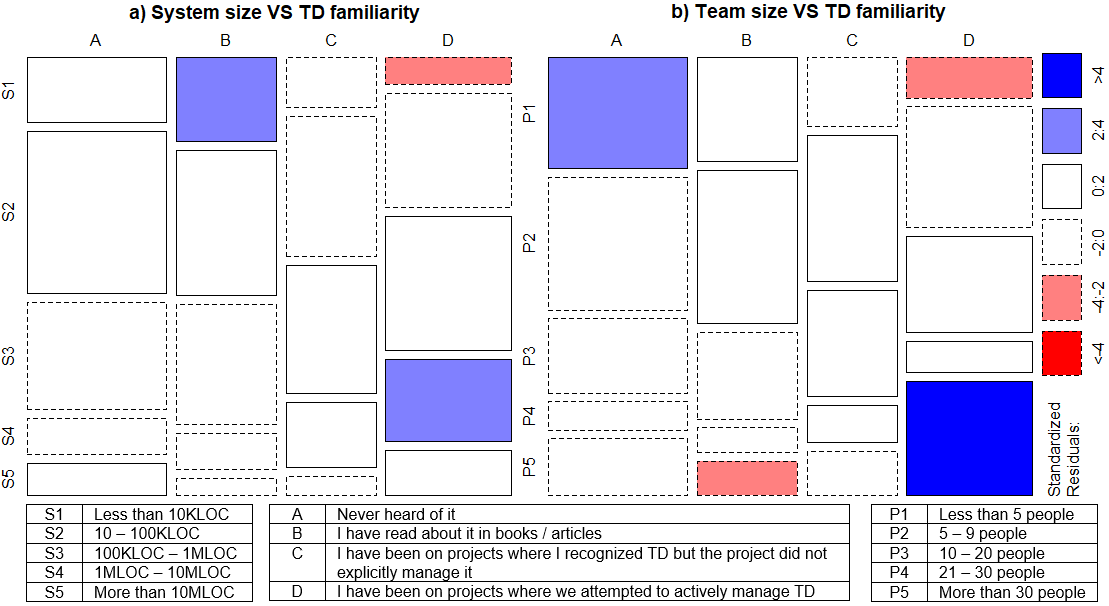}
  \caption{The mosaic plot for system and team size versus TD familiarity (Q9)}
  \label{fig.mosaic.SystemTeamSizeVSQ9}
\end{figure}

\paragraph{System size versus TD familiarity}
Figure \ref{fig.mosaic.SystemTeamSizeVSQ9}.a depicts a mosaic plot of prevalence of the TD concept in industry, with respect to system size. The analysis of the plot in the figure shows that practitioners that work on systems that have less than 10KLOC are more likely to have only theoretical knowledge of TD and are less likely to have practical experience with managing TD. 
On the other hand, practitioners that work on systems that have between 1 and 10MLOC are more likely to have experience in TD management.

\paragraph{Team size versus TD familiarity}

Looking into teams size versus TD familiarity (Figure \ref{fig.mosaic.SystemTeamSizeVSQ9}.b) it can be observed that practitioners who work in large teams, more than 30 people, are the most likely to have practical experience with managing TD, and are less likely to only have theoretical knowledge of TD. On the other side, practitioners that work in small teams, less than 5 people, are more likely to have never heard of the TD concept and less likely to have practical experience in managing TD.

\begin{figure}
  \centering
  \includegraphics[width=12cm, height=6.5cm, frame]{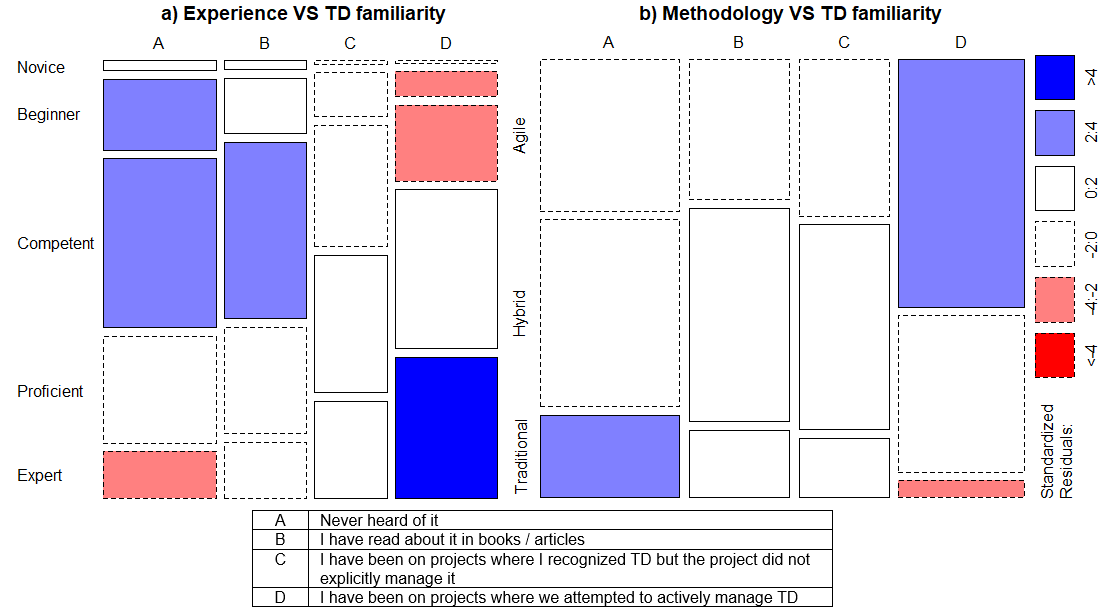}
  \caption{The mosaic plot for methodology and practitioner experience versus TD familiarity (Q9)}
  \label{fig.mosaic.ExperienceMethodologyVSQ9}
\end{figure}

\paragraph{Experience versus TD familiarity}
Figure \ref{fig.mosaic.ExperienceMethodologyVSQ9}.a depicts the prevalence of the TD concept in industry, with respect to practitioner experience. Practitioners that are beginners are more likely to have never heard of the TD concept and also less likely to have practical experience in TD management. Practitioners that classify as competent are more likely to have never heard of the TD concept, and also equally likely to have theoretical knowledge on TD. Also, like beginners, they are less likely to have practical experience in managing TD. Expert practitioners are less likely to have never heard of TD and most likely to have practical experience in TD management.

\begin{finding}
\label{fin.Finding3}
As the system size or company size or team membership count or level of expertise increases, the likelihood that IT practitioners have experience with TD management increases as well. 
\end{finding}

\begin{figure}
  \centering
  \includegraphics[width=9.5cm, height=8cm, frame]{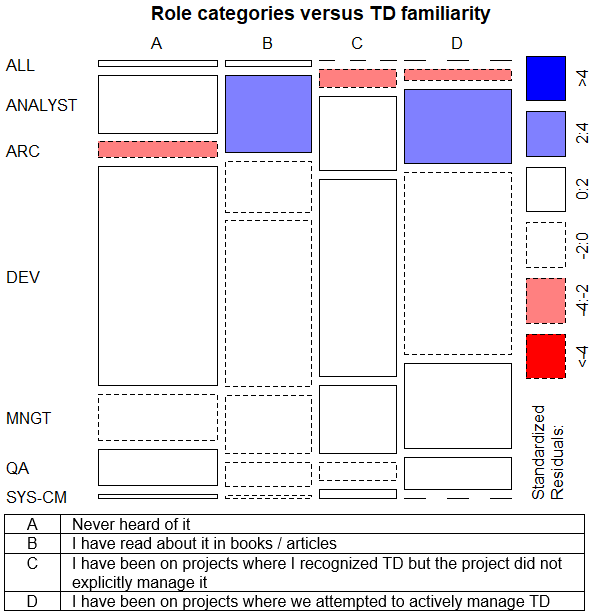}
  \caption{The mosaic plot for role categories versus TD familiarity (Q9)}
  \label{fig.mosaic.RoleVSQ9}
\end{figure}

\paragraph{Role versus TD familiarity}
\label{rq1.mosaic.roleVSQ9}
Figure \ref{fig.mosaic.RoleVSQ9} depicts a mosaic plot of prevalence of the TD concept in industry, with respect to practitioner roles. For this analysis practitioner roles were categorized in order to generate a better illustration of the association between roles and TD familiarity. The categories are presented in Table \ref{tab.RoleCategories}. After categorizing participant roles the $\chi^2$ test was run again and resulted in the following: $\chi^2$ = 73.845, df = 18, p-value = 9.997e-09. As the results prior to role categorization these results indicated that there indeed is a statistically significant association in regards to TD familiarity.

\begin{table}[h]
    \caption{Categories of participant roles}
    \label{tab.RoleCategories}
    \centering
    \scriptsize 
    \begin{tabular}{ll}
        \toprule
        Category & Roles assigned to category\\
        \midrule
        {DEV}       &{Developer}  \\
        {MNGT}      &{Project leader/project manager, Quality analyst manager}  \\
        {ARC}       &{Software architect}  \\
        {QA}        &{Tester, Quality analyst}  \\
        {ANALYST}   &\makecell[l]{Process analyst, Requirement analyst, \\ Database administrator/analyst, Business analyst}  \\
        {ALL}       &{Performs multiple functions}  \\
        {SYS-CM}    &{Configuration manager, Infrastructure analyst, Infrastructure manager} \\
        \bottomrule
    \end{tabular}
\end{table}

Analysis of Figure \ref{fig.mosaic.RoleVSQ9} shows that if practitioners fall into the \textit{Analyst} category they are more likely to have theoretical knowledge of TD but are less likely to have practical experience in TD identification or management. Practitioners that are software architects are less likely to have never heard of the TD concept and more likely to actually have some practical experience in TD management. Practitioners that are developers seem to be equally likely to have never heard of the TD concept, have theoretical knowledge about TD or have some practical experience in TD identification and management. The same can be said for practitioners that fall into the \textit{MGMT} and \textit{QA} categories. The \textit{ALL} and \textit{SYS-CM} seem to lean this way also, but the sample is insufficiently large to make any assumptions.

\begin{finding}
\label{fin.Finding4}
Practitioners working as software architects are more likely to have practical experience in TD management, whereas analysts, despite their theoretical knowledge about TD, are more likely to have never participated in a project where TD was recognized or managed.  
\end{finding}

\paragraph{Methodology versus TD familiarity}
Figure \ref{fig.mosaic.ExperienceMethodologyVSQ9}.b depicts the prevalence of the TD concept in industry, with respect to the used development approach. The data shows that traditional development approaches are more likely to have practitioners working with them that have never heard of the TD concept and less likely to have practical experience in TD management. On the other hand, practitioners using agile approaches are more likely to have practical experience in TD management.

\begin{finding}
\label{fin.Finding5}
Practitioners using non-agile like approaches, i.e. traditional approaches, are more likely to have never heard of the TD concept and less likely to have any experiences with TD management.
\end{finding}

\subsection{Technical Debt Causes and Effects}
\label{sec.Results.TDCausesEffects}

The participants gave their insight into the causes and effects of TD, in the context of the exemplar project that they specified. The analysis of causes and effects of TD consisted of the: (a) conceptualization of TD causes, (b) identification of the most common causes and effects of TD, and (c) identification of associations between the TD types (Section \ref{sec.CategoriesTDTypes}) on one side, and causes and effects on the other. Most common causes and effects were identified using coding and categorization and the results were counted and charted. The associations between the TD types, and causes and effects were identified using the $\chi^2$ test and contingency tables, while for the presentation of these associations, the mosaic plots were used. The analysis results contribute to answers to RQ2 and RQ3.

\subsubsection{Conceptualization of TD causes}
\label{sec.Rq2.conceptualizatitonTDcauses}
 The TD causes were extracted using the participants answers to survey questions Q16-Q19. Further analysis of these causes led to the derivation of eight categories. These categories, ordered by occurrence frequency and with exemplary causes are: 

 \begin{enumerate}
    \item \textit{Planning and management (725)}: Multitasking different clients and pro\-jects, misconduct, poor leadership.
    \item \textit{Development issues (473)}: Lack of quality, legacy system, legacy artifacts.
    \item \textit{Methodology (389)}: Tests not performed, poor task distribution.
    \item \textit{Lack of knowledge (311)}: Lack of experience, information.
    \item \textit{People (208)}: Overconfidence, Non-sharing of knowledge, lack of team communication. 
    \item \textit{Organizational (133)}: Lack of training, lack of priority for the project.
    \item \textit{External factors (119)}: Payment dynamics, discontinued component.
    \item \textit{Infrastructure (42)}: Inadequate choice of technology, tool, platform, required infrastructure unavailable. 
\end{enumerate}

\begin{figure}[h]
  \centering
  \includegraphics[width=\textwidth, frame]{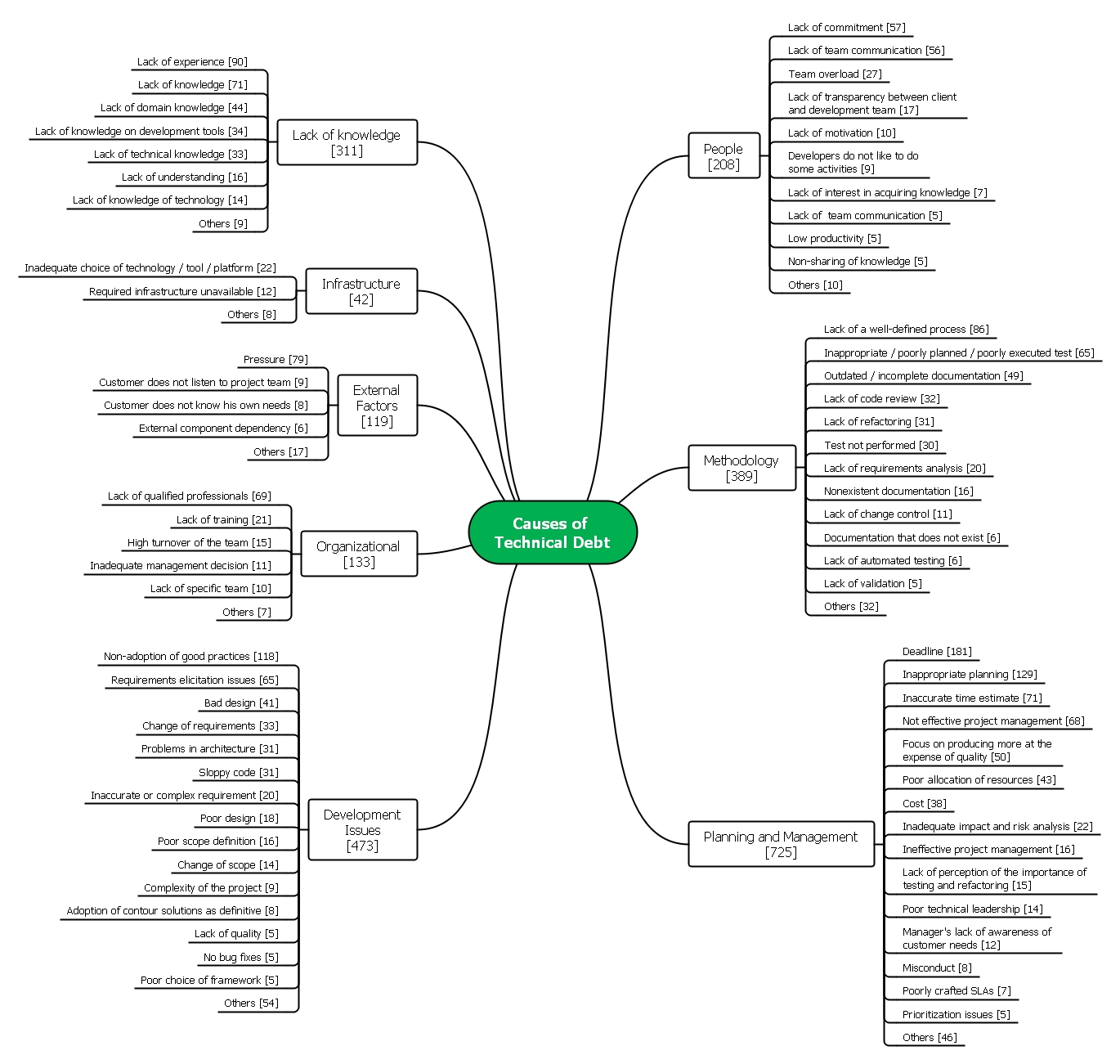}
  \caption{Mind map of TD causes. Causes with the frequency bellow 5 are omitted and grouped together as the \textit{Other} item for each category in the map. High-resolution map with all causes is available at \url{https://bit.ly/3gjt0Bc}}
   \label{fig.mindmapcauses}
\end{figure}

The identified causes, derived categories, and their relationship, as well as their frequencies, represents the conceptualization of TD causes. 
This conceptualization is presented in a from of mind map on Figure \ref{fig.mindmapcauses}. The mind map reveals that the two dominant categories or group of causes are the \textit{Planning and management} and \textit{Development issues}, and the causes that most frequently occurred in these categories.

\subsubsection{The most common causes of TD}
\label{sec.Rq2.Top10Causes}

 The most dominant cause of TD by far is the \textit{Deadline} cause, indicating that developers mostly incur debt because of tight deadlines that are set, and using quick solutions that are not thought through thoroughly. \textit{Inappropriate planning} and \textit{not effective project management} also rank high and indicate a problem with leading the project. These causes are somewhat tied together on the count of them both referencing project management, one from the planning aspect and the other from a more operational aspect of leading the project in real time. Other significant causes are: \textit{Focus on producing more at the expense of quality}, \textit{Lack of experience}, \textit{Pressure}, \textit{Non-adoption of good practices}, \textit{Inaccurate time estimate}, \textit{Lack of qualified professionals}, \textit{Lack of a well-defined process}.

\begin{figure}
  \centering
  \includegraphics[width=12cm, height=5.5cm, frame]{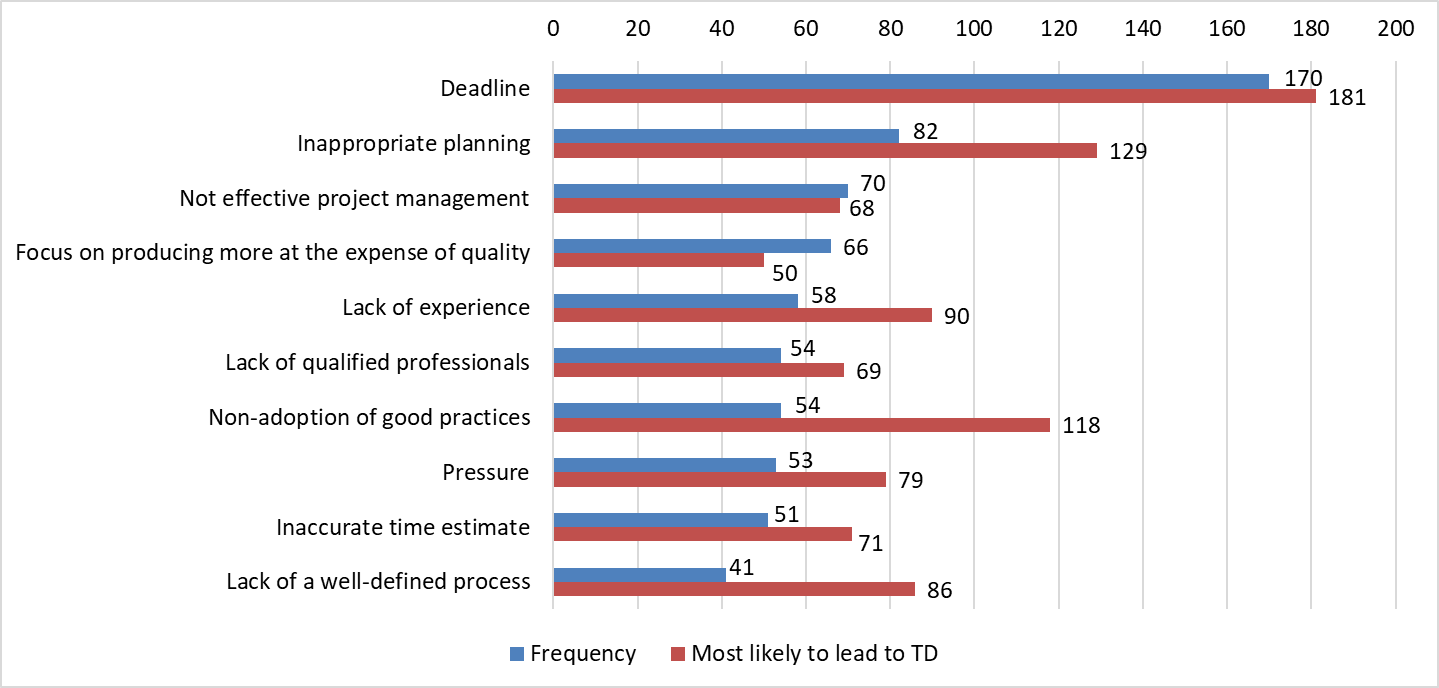}
  \caption{Top 10 most cited causes of TD}
   \label{fig.Top10Causes}
\end{figure}

Further analysis of Figure \ref{fig.Top10Causes} reveals that the likelihood of some causes leading to TD was assessed by participants as more likely, even though they had a lower citing frequency. \textit{Non-adoption of good practices} seems to be right behind \textit{Deadline} and \textit{Inappropriate planning} regarding the possibility of it leading to TD. Also, \textit{Lack of experience}, \textit{Lack of a well-defined process} and \textit{Pressure} follow, indicating that they must not be ignored when analyzing possible causes of TD.

\begin{finding}
\textit{Deadlines}, \textit{non-adoption of good practices}, \textit{lack of experience} and \textit{pressure} are the most likely causes why IT practitioners incur TD. These causes are often a manifestation of other, indirect, causes which are most likely inappropriate planning, lack of well-defined processes or non-effective project management.
\end{finding}

\subsubsection{Categories of Causes and TD Types}
\label{sec.CategoriesTDTypes}
\begin{figure}
  \centering
  \includegraphics[width=12cm, height=6.5cm, frame]{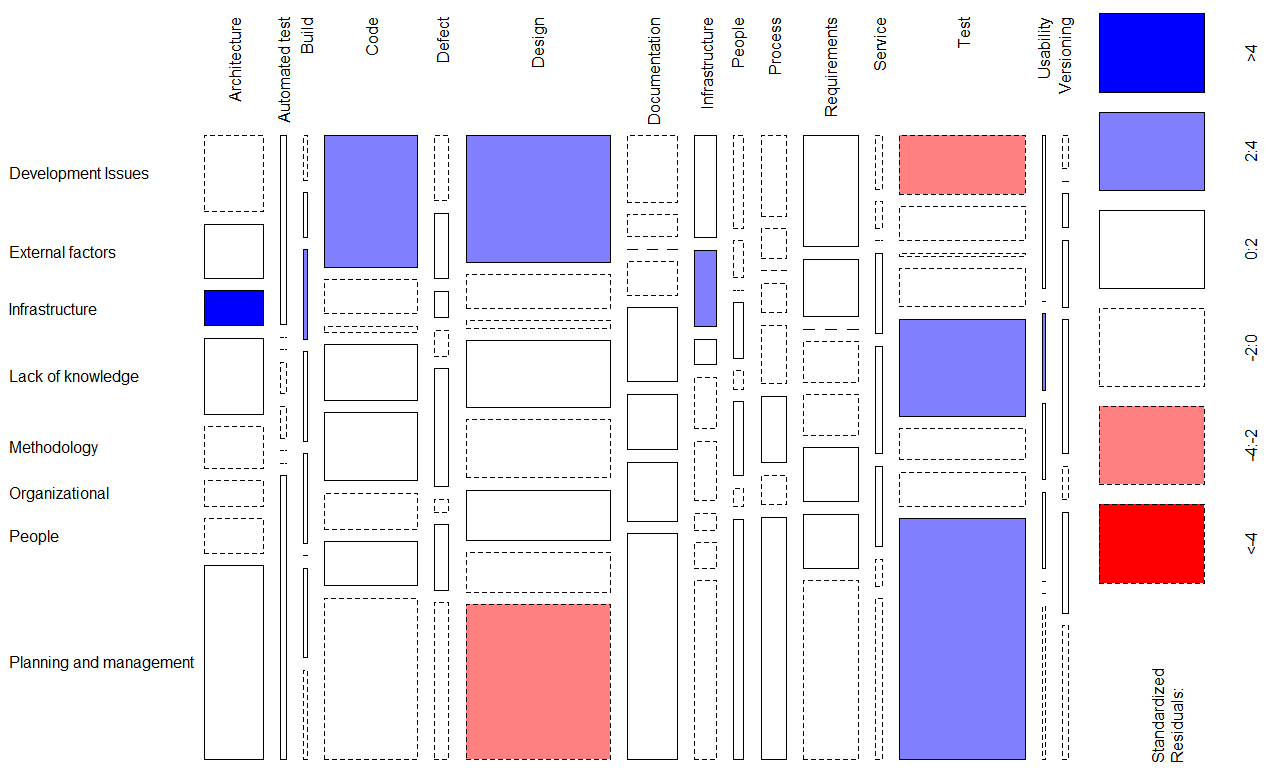}
  \caption{The mosaic plot for categories of causes versus TD Types}
   \label{fig.mosaic.CausesCatVSTDType}
\end{figure}

As presented in Section \ref{sec.Rq2.conceptualizatitonTDcauses}, the causes were categorized into 8 categories. The $\chi^2$ test calculated to assess the relationship between TD categories and types resulted in the following: $\chi^2$ = 214.59, df = 98, p-value = 1.067e-10. The p-value indicated on a statistically significant relationship between categories of TD causes and TD types. Figure \ref{fig.mosaic.CausesCatVSTDType} shows the mosaic plot.

The analysis of Figure \ref{fig.mosaic.CausesCatVSTDType} shows that if TD causes fall into the \textit{infrastructure} category they are most likely to lead to \textit{architecture debt} and a bit less likely to lead to \textit{build debt} or \textit{usability debt}.  Causes that are external in nature are more likely to lead to \textit{infrastructure debt}, or issues with the way the organization itself operates (e.g. using outdated tools, postponing updates). Causes that have to do with the development approach (methodology) are more likely to lead to \textit{test debt}.

Section \ref{sec.Rq2.conceptualizatitonTDcauses} revealed that \textit{planning and management} and \textit{development issues} are the two dominant categories of causes.
Causes categorized as \textit{planning and management} are less likely to lead to \textit{design debt} but more likely to lead to \textit{test debt}. Causes that are categorized as \textit{development issues} are more likely to lead either to \textit{code debt} or \textit{design debt} but are less likely to lead to testing related issues or \textit{test debt}. \textit{Test debt} is therefore most likely caused by \textit{planning and management} group of causes, while the \textit{design debt} is caused by \textit{development issues} causes.

\begin{finding}
\label{fin.Finding7}
\textit{Infrastructure} causes are most likely to lead to architecture debt, and a bit less likely to lead to build or usability debt.
\textit{Development issue} causes are more likely to lead either to code or design debt but are less likely to lead to test debt.
\textit{External factor} causes are more likely to lead to infrastructure debt.
\textit{Planning and management, and methodology} causes are more likely to lead to test debt. 
\end{finding}

\subsubsection{Conceptualization of TD effects}
\label{sec.RQ3.conceputalizationEffects}

The TD causes were extracted using the participants answers to survey question Q20 and Q21. The analysis of Q21 gave insight into effects TD have on development process and these effects were categorized into six categories. These categories, their frequencies and exemplary effects are: 
\begin{enumerate}
    \item \textit{Planning and management (467)}: Increased costs, increased efforts.
    \item \textit{External quality (338)}: Low quality, low performance, low maintainability.
    \item \textit{Internal quality (308)}: Poor code readability, increased amount of maintenance activities.
    \item \textit{People (395)}: Customer or user dissatisfaction, Lack of knowledge.
    \item \textit{Development issues (282)}: Low code reuse, low tractability.
    \item \textit{Organizational (162)}: Financial loss, Impaired company image.
\end{enumerate}

\begin{figure}
  \centering
  \includegraphics[width=\textwidth,  keepaspectratio, frame]{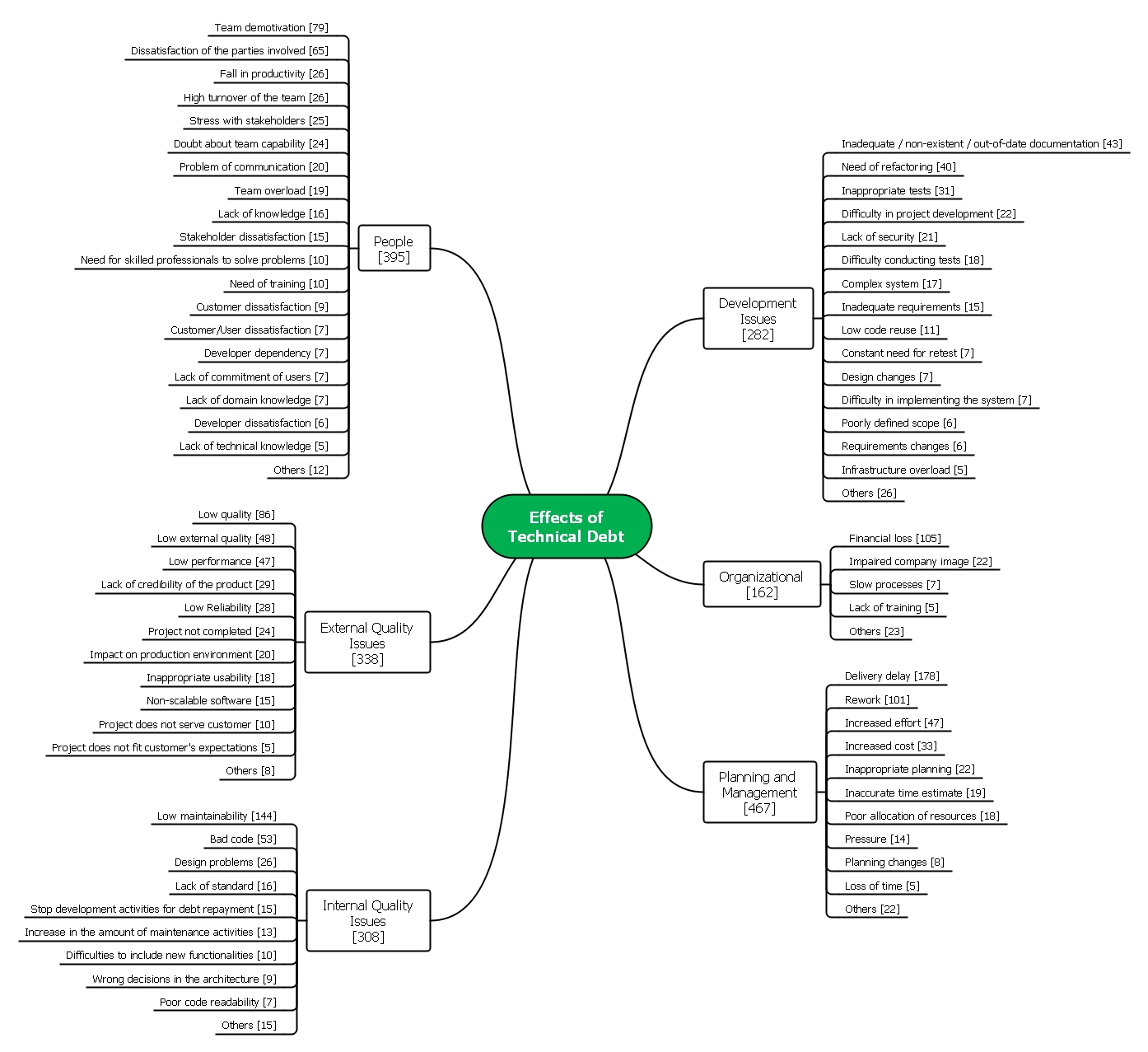}
  \caption{Mind map of TD effects. Effects with the frequency bellow 5 are omitted and grouped together as the \textit{Other} item for each category in the map. High-resolution map with all effects is available at \url{https://bit.ly/3gtjVVo}}
   \label{fig.mindmapeffects}
\end{figure}

As with TD causes, effects and their categories are arranged in mind map, and this map is presented in Figure \ref{fig.mindmapeffects}. Mind map reveal that Planning and management and External quality issues are the most dominant group of TD effects. 

\subsubsection{The most common effects of TD}
\label{sec.RQ3.Top10Effects}

\begin{figure}
  \centering
  \includegraphics[width=12cm, height=5.5cm, frame]{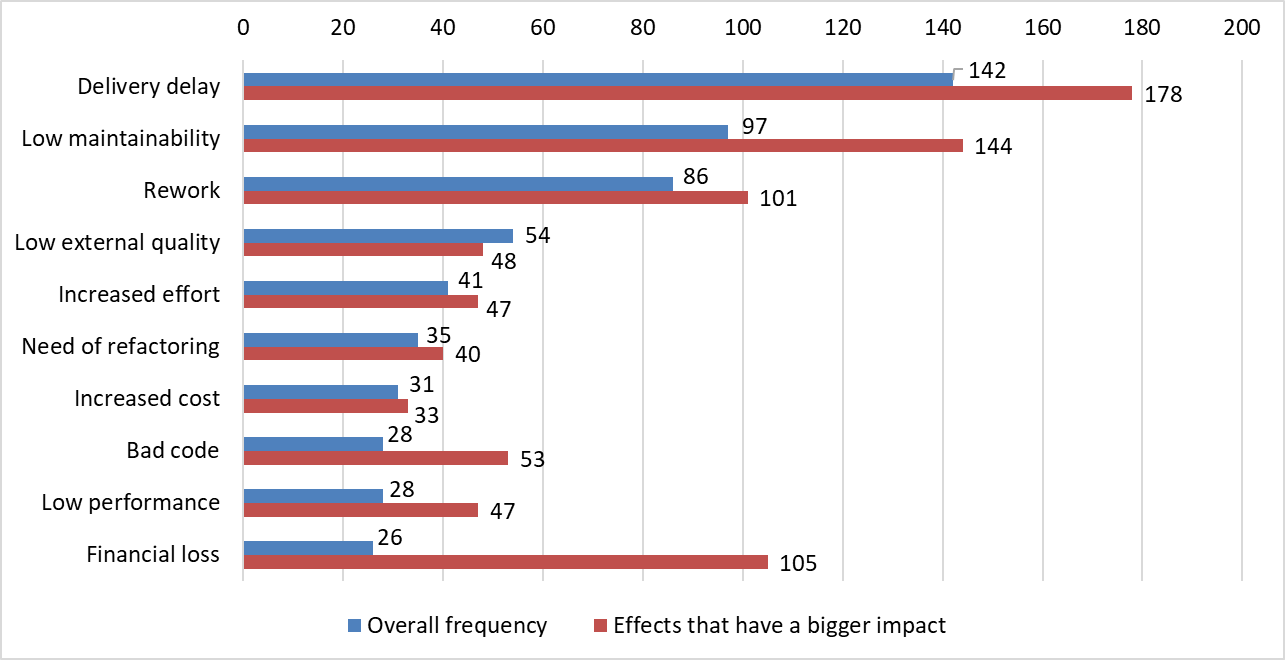}
  \caption{Top 10 most cited effects of TD}
   \label{fig.Top10Effects}
\end{figure}

 The most dominant effect of TD is \textit{Delivery delay}, as presented in Figure \ref{fig.Top10Effects}. This effect is also the one with the biggest impact. \textit{Low maintainability} and \textit{Rework} are also highly cited effects of TD, as well as the effects with a significant impact. Other significant effects are: \textit{Low external quality}, \textit{Increased effort}, \textit{Need of refactoring}, \textit{Increased cost}, \textit{Bad code}, \textit{Low performance} and \textit{Financial loss}.

One other significant insight that can be obtained from Figure \ref{fig.Top10Effects} is that \textit{Financial loss} is cited last in the top 10 effects list, but is categorized as the effect with the third highest impact. A conjecture can be made that this is mostly viewed by the management on the count of this effect being directly related to project, and down the line corporate, financial state.

\begin{finding}
The presence of TD most likely leads to delivery delays, low maintainability and generates the need for rework. These effect can also lead to longer-term effect, most likely financial loss.  
\end{finding}

\subsubsection{TD Types and Categories of Effects}
\label{sec.RQ3.typesandsymptoms}

\begin{figure}
  \centering
  \includegraphics[width=12cm, height=6.5cm, frame]{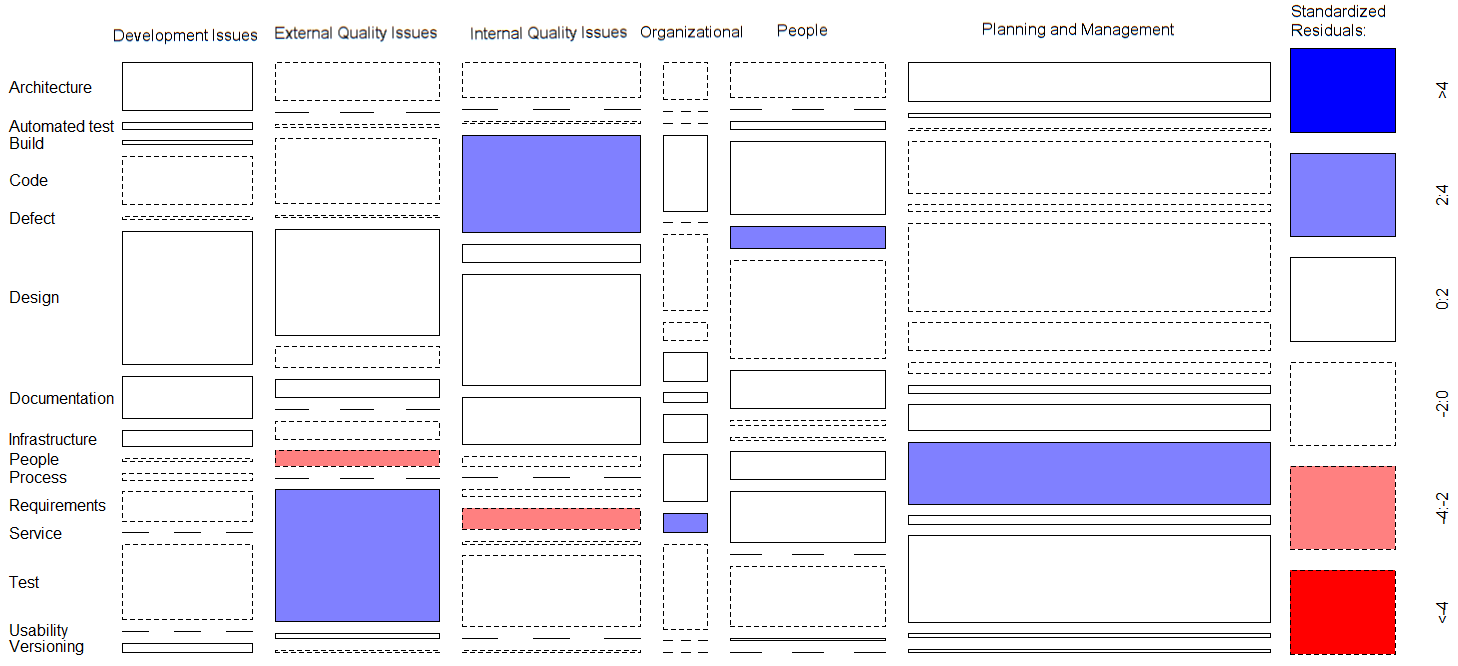}
  \caption{The mosaic plot for TD Types versus categories of TD effects}
   \label{fig.mosaic.TDTypeVSEffectsCat}
\end{figure}

As it was the case with TD causes, the effects of TD were categorized in six categories. The $\chi^2$ test that assesses the relationship between TD effects and TD types, resulted in the following: $\chi^2$ = 131.14, df = 70, p-value = 1.336e-05. The p-value indicated that there indeed exists a statistically significant relationship between TD types and categories of TD effects. Figure \ref{fig.mosaic.TDTypeVSEffectsCat} shows the mosaic plot that was generated to visualize this relationship and to further analyze it.

Figure \ref{fig.mosaic.TDTypeVSEffectsCat} shows that \textit{code debt} most likely leads to issues that can be categorized as \textit{internal quality issues}. 
This is somewhat expected on the count of the code being owned by the company and any issues that arise from it can be treated as internal to that company. 
\textit{Defect debt} represents defects that are discovered by testing activities but not fixed or corrected. This type of debt is most likely linked to effects that have impact on the people (e.g. developers being demotivated by some product defects). Issues that arise when the selection of web services is not aligned with requirements of the system, or \textit{service debt} is more likely to be related to effects that can be categorized as \textit{organizational}.  
Section \ref{sec.RQ3.conceputalizationEffects} revealed that the two most dominant groups of TD effects are the \textit{Planning and Management} and \textit{External Quality Issues}. Debt that refers to issues that arise from the trade off between agreed upon requirements and the ones that are actually implemented, or \textit{requirements debt}, is more likely to be linked to effects that resonate with the management level and fall into the \textit{planning and management category}, but are less likely to be linked to \textit{external or internal quality issues}. On the other hand, debt that can be categorized as \textit{test debt} is more likely to be related to effects that are external, or fall into the \textit{External quality issues} category. \textit{Requirements debt} will therefore likely lead towards the effects from \textit{planning and management category}, while the \textit{test debt} most likely leads to effects from \textit{external quality issues} category.

\begin{finding}
The code and test types of TD are likely to lead to issues with internal and external quality, respectively.
While, requirements debt is likely to lead to planning and management issues. 
Furthermore, defect debt is associated with people related issues.
\end{finding}

\section{Discussion}
\label{sec.Discussion}

This chapter discusses the most significant findings of the study. Section \ref{sec.Discussion.Prevalence} discusses the findings relevant to the prevalence of the TD concept, while Section \ref{sec.Discussion.Causes} and Section \ref{sec.Discussion.Effects} focus respectively on the findings that emerged from the analysis of TD causes and effects.

\subsection{The prevalence of TD}
\label{sec.Discussion.Prevalence}

Looking at the prevalence of TD, the study findings indicate that the industry is generally familiar with the TD concept with 78\% of participants confirming that they understand or agree with McConnell's definition of TD (Finding \ref{fin.Finding1Prevalence}). On the other hand, even though the TD familiarity percentage is relatively high, TD identification and management is somewhat less present in the IT industry with more than half of the participants of this study (53\%) indicating that they did not work on projects that attempted to identify or monitor TD (Finding \ref{fin.Finding1Prevalence}). These results suggest that the majority of the industry is at least familiar with the concept of TD, while a certain percentage has knowledge about TD but has no concrete experience with it. Could this be on the count of organizations themselves not willing to track TD, or is TD knowledge lacking on the management level? Either way, the results show that there is a certain percentage of practitioners that are familiar with TD and could possibly help mitigate its impacts but whose knowledge is not being utilized.

A study by Silva et al. \cite{silva2019taste} set a similar research question to this studies RQ1, namely they strove to see if there is a consensus on the perception of TD among software practitioners in the Brazilian software organizations. They analyzed responses from 40 participants and showed that 60\% of them claimed to be aware of the TD concept. Generalizing to this study this corresponds well to the majority of participants claiming that they are familiar with the TD concept.

These participants came from three large companies and the study found that they are generally familiar with the concept of TD. This results coincides well with this studies Finding \ref{fin.Finding3} that states that as the company size increases so does the likelihood that practitioners are familiar with the concept of TD, and also are more likely to have some practical experience in TD identification or management.

An interesting result is the one expressed in Finding \ref{fin.Finding2CountryVSPrevalence} that states that there is a dependency between country and the familiarity with the TD concept. To the best of our knowledge no other study of international character has detected this dependency. The authors' opinion is that two conjectures can be set regarding the explanation of this finding. The first one is that this is primarily because of the divers education system (regarding IT) that are in place in different countries, some of which cover the topic of TD in more detail, while others cover the topic more loosely or not at all. This has the most significant impact on junior developers and their familiarity and perception of the TD concept. The second conjecture is regarding the maturity of the national IT industry. Some IT industries are less mature than others, maybe they are still young, and it is in these industries that practitioners are less familiar with TD. Figure \ref{fig.mosaic.CountryCompanySizeVSQ9}.a hints at this with it results, showing that if practitioners come from Serbia, that has a relatively young IT industry that is still developing, they are more likely to have never heard of the TD concept. 
The variations in national data can be also caused by cultural characteristics, a further research is needed to investigate the cultural aspects of TD.

Shifting the focus to architectural TD, the literature shows that there were studies like the study from Martini et al. \cite{martiniEUROMICRO2014} and Martini and Bosch \cite{martiniATD2017} that focused on this particular type of TD. This study did not find that the architecture as an artifact is a top TD cause. Section \ref{sec.Demographics.TDTypes} shows that \textit{architectural debt} came in fourth with 10.70\% share in total TD types detected from participants exemplar projects, although Finding \ref{fin.Finding4} states that software architects are more likely to have practical experience in TD management. Being that software architects are not only entrusted with and responsible for designing the system, but are also responsible for the software as a whole meeting all agreed upon requirements and standards, it is only logical that they are, among some other roles, most interested in TD.

When it comes to the methodology that is being used, practitioners that use traditional approaches are more likely to have never heard of the TD concept and less likely to have some experience in TD management (Finding \ref{fin.Finding5}). It can be argued that this is somewhat expected on the count of more traditional approaches being less present in software development nowadays, while Agile approaches, or even hybrid approaches being more dominant.

Rocha J. C. et al. \cite{rocha2017understanding}, among other things, reported on knowledge on TD from the perspective of 74 Brazilian practitioners. They stated that 25.7\% had low knowledge on TD, 36.5\% a medium grasp on TD and 25.7\% a high understanding of TD. Figure \ref{fig.mosaic.CountryCompanySizeVSQ9} a) shows that this study also points to Brazilian practitioners being more likely to have theoretical knowledge on TD.

Ernst et al. \cite{ernst2015measure} in their study found that the TD metaphor is widely accepted within large organizations. This study did not find any statistically significant relationship between company size and the familiarity of TD, only that it is less likely that practitioners that come from small companies have actively managed TD on some projects within that company \ref{fig.mosaic.CountryCompanySizeVSQ9}.b).

\subsection{Top Causes of TD}
\label{sec.Discussion.Causes}
The top most cited cause of TD was the \textit{Deadline} cause, with the citing frequency of 170. Besides deadlines some of the most cited causes of TD that emerged were causes such as \textit{Inappropriate planning}, \textit{Not effective project management} and \textit{Focus on producing more at the expense of quality}. An interesting cause was the \textit{Non-adoption of good practices} cause, that came in the seventh place in the top 10 causes list, but is the third cause that is, according to study participants, most likely to lead to TD, which makes this cause worthy of considering even before some that have a higher cite frequency. Other causes that fall into the top 10 causes of TD list can be viewed in Figure \ref{fig.Top10Causes}.

When analyzing the likelihood of a cause leading to TD, the \textit{Deadline} cause was cited a quarter of the times (25\%) in the \textit{Planning and Management} category, as the most likely cause to lead to TD \ref{fig.mindmapcauses}. It also took up 26\% of citations of all top cited causes in remaining 7 categories. In the same category, other causes emerged, like the \textit{Inappropriate planning} cause, that had 18\% of that categories citations, the \textit{Inaccurate time estimate} cause with 10\% of citations, the \textit{Not effective project management} cause with 9\% of citations, and the \textit{Focus on producing more at the expense of quality} with 7\%. These causes take up 69\% of citations in the \textit{Planning and Management} category and take up more than half (53\%) of citations in the top 10 causes list regarding the likelihood of leading to TD \ref{fig.Top10Causes}. This data points to \textit{Planning and Management} as the potential reservoir of potential TD causes.

The \textit{Deadline} as a cause was recognized by Yli-Huumo et al. \cite{YliHuumoPROFES2014} in their study as lack of time given for development. They also recognized pressure put on the development team as a cause of TD which can coincide with this studies \textit{Pressure} cause that was cited 53 times and recognized by practitioners as relatively highly likely to lead to TD.

The study that was conducted by Codabux and Williams \cite{codabux2013managing} indicated that managerial decisions are among the main causes of TD. These decisions can result in inappropriate planning, not effective project management or even a lack of a well defined process, that was also detected by this study as a highly cited cause. This study confirmed that management decisions, at different management levels (e.g. project management, organization management), can in deed be a cause of TD. The study in \cite{codabux2013managing} also recognized resource constraints as a cause of TD. This was also confirmed by this study in the sense that \textit{Lack of qualified professionals} was detected as a highly cited cause of TD with a frequency of 54.

One of the largest studies on TD that was conducted by Ernst et al. \cite{ernst2015measure} came to the conclusion that immature choices done in architecture are a key cause of TD. The \textit{Inappropriate planning} cause identified by this study can be a result of poorly made decisions in architecture that lead to TD.

The study conducted by Martini et al. \cite{martiniEUROMICRO2014} focused on architecture debt and strove to find the causes for the accumulation of architectural TD and also to investigate the trends in the accumulation and recovery of architectural TD. The study found that business factors, design and architecture documentation, reuse of legacy/third party/open source, parallel development, effects uncertainty, non-completed refactoring, technology evolution, and the human factor can be the likely causes. This study builds on this and finds that \textit{Infrastructure} causes, or inadequate technology, tool, platform choices, or an unavailable infrastructure, are most likely to lead to architecture debt. Besides this, the data shows that code or design debt will most likely be triggered by some \textit{Development issues} cause, that \textit{External factor} causes most likely lead to infrastructure debt and \textit{Planning and management} and \textit{Methodology} causes most likely lead to test debt.

The identification of several causes of TD that were mentioned in earlier studies by other authors can possibly indicate that a trend is slowly emerging for some causes that are most likely to trigger the existence of TD items, but more research is needed to derive some conclusions that are generalizable.


Ernst et al. \cite{ernst2015measure} in their study, showed that architecture decisions were the main source of TD, while this study adds the finding that several other causes that can be classified as \textit{Infrastructure} causes can lead to Architecture debt (\ref{fig.mosaic.CausesCatVSTDType}). They also found \textit{Overly complex code} to be a significant cause. The debt that occurs in this context can be viewed as \textit{Code} and this study found that causes categorized as \textit{Development Issues} can lead to this particular debt type (\ref{fig.mosaic.CausesCatVSTDType}). Architecture as a potential leading cause of TD is also the found by Holvitie et al. \cite{holvitie2018technical}. They also found that inadequate tests seem to be a cause of TD, whereas this study showed that causes categorized as \textit{Planning and Management} and \textit{Methodology} can lead to Test debt (\ref{fig.mosaic.CausesCatVSTDType}).

\subsection{Top Effects of TD}
\label{sec.Discussion.Effects}
The top most cited effect of TD was the \textit{Delivery delay} effect, cited 142 times, meaning that in most cases practitioners saw that because of debt items they were not able to reach an agreed upon deadline for delivery. This is an interesting find in comparison to the top TD cause being the \textit{deadlines} that are defined. A vicious circle seems to form here on the count of practitioners rushing to meet the deadlines and introducing TD in the project that in turn later delays the delivery of the project. Besides this, some of the other highly cited effects that practitioners named were effects like \textit{Low maintainability}, \textit{Rework} and \textit{Low external quality}. An interesting effect in the top 10 list was the \textit{Financial loss} effect, that came in 10th regarding citing frequency, but was the third effect regarding the significance of its impact. This is somewhat expected because this effect speaks about company finances, and TD leading to financial loss can have serious repercussions on the company as a whole.

When analyzing what effects have a bigger impact it seems that \textit{Delivery delay}, \textit{Low maitainability}, \textit{Financial loss}, \textit{Rework} are the ones with the biggest impact \ref{fig.Top10Effects}. The \textit{Delivery delay} cause comes from the \textit{Planning and Management} category and takes up 38\% of effects cited in that category. On the other hand, \textit{Low maitainability} is placed in the \textit{Internal Quality Issues} category and also takes up 47\% of effects in that category. The \textit{Financial loss} effect is probably felt the most by the management of the company and is categorized in the \textit{Organizational} category and it is taking up the majority of that category with 65\% of all cited effects. The \textit{Rework} effect comes also from the \textit{Planning and Management} category with a share of 22\%. These effects combined take up 66\% of effects that have the biggest impact, and should be taken into consideration first when analyzing TD.

In the Codabux and Williams \cite{codabux2013managing} study the results showed that starting the project from scratch was the main effect felt by practitioners, while this studies participants listed \textit{rework} as the third most common effect of TD. Yli-Huumo et al. \cite{YliHuumoPROFES2014} extra working hours and errors and bugs were among the main effects of TD. This study confirms this by identifying \textit{increased effort} and \textit{bad code} as effects that are among the most cited by study participants. This studies findings for top effects also complement the study by Yli-Huumo et al. \cite{yli2015benefits} that found, among others, decreased code maintainability, major refactoring as the emerging effects of TD. Similarly the study by Li et al. \cite{li2015systematic} also found that maintainability is one quality attribute that is compromised because of the presence of debt items.

This studies findings confirm some of the earlier study finding regarding the effects of TD, but also build upon the work by introducing some new effects that are recognized by practitioners as relevant.

Besker et al. \cite{besker2018technical} argued that additional time that is wasted on a project is spent in performing additional tests, and thus makes an important effect of TD to note. This study showed that all effects that can be categorized as \textit{External Quality Issues} are likely to lead to Test debt Finding \ref{fig.mosaic.TDTypeVSEffectsCat}. The authors of \cite{besker2018technical} also mention additional source code analysis and refactoring as most common effects of TD. These effects can be categorized as \textit{Development Issues} for which this study did not find any statistically significant relationship with any TD type Finding \ref{fig.mosaic.TDTypeVSEffectsCat}. Yli-Huumo et al. \cite{yli2015benefits} mentioned \textit{decreased code maintainability} as one of the most dominant consequences for quick workarounds, while this study categorizes low maintainability as an \textit{Internal Quality Issue} and this TD effect category is likely to lead to Code debt.

\subsection{InsighTD Dataset}
\label{sec.InsighTDDataset}

In this publication we reported the InsighTD dataset with respect to prevalence, causes, and effects data. The given analysis aids better description of the dataset, i.e. it supports more exploratory analysis, rather than a specific data analysis for preset hypotheses. However, the intention is to allow formulations of such hypotheses in future studies based on this report. 

Our intention is to present a report of InsighTD dataset that will allow for versatile usage. For example, from data on causes it can be seen that causes characterized as \textit{planning and management }are far more common than other causes (Figure \ref{fig.mindmapcauses} in Section \ref{sec.Results.TDCausesEffects}). Furthermore, data also suggest that in presence of planning and management causes testing practices will suffer the most, i.e. it will result with the injection of a \textit{test debt} (Figure \ref{fig.mosaic.CausesCatVSTDType}). So, reading segments of this report provides different insights that can motivate further investigations.

As one of the objectives of InsighTD is to increase the dataset by adding new replication countries, the dataset will change, i.e. grow, over time. An interested reader can find an updated version of the InsighTD analysis report on the following link\footnote{InsighTD supplementary analysis report. Version 1.0. (Date: 10.06.2021) \url{https://bit.ly/3xeU4Hu}} where all additional analysis of InsighTD data are included. 

\section{Trustworthiness of the Study}
\label{sec.Trustworthiness}

The assessment of the trustworthiness of the study depends on its underlying paradigmatic or philosophical stances. 
This study adopted the constructivism as a paradigmatic stance \cite{Creswell2008}.  Constructivism's assumptions were seen as most acceptable considering the study, its objective and the stated RQs. To argument this decision, constructivists' assumptions, proposed in Guba and Lincon \cite{Lincoln2011} will be presented alongside the properties of this study and the InsighTD project. The arguments are as follows: (a) For constructivism, the inquiry goal is the understanding of the phenomenon. 

The goal of InsighTD is to be achieved by understanding the practitioners' familiarity with TD and their reactions when TD is present, the causes of TD and effects it has on software development process. (b) For constructivists, knowledge is inseparable from humans and their context. In InsighTD, knowledge is generated using IT industry practitioners' experiences with TD. (c) Constructivists focus on distinct groups of people and seek to find their local truths. InsighTD builds on findings from national, that is on local, findings and aggregates them to derive the new knowledge. (d) Constructivists typically rely on qualitative methods for data collection and analysis. InsighTD survey includes open questions to collect rich qualitative data and this data is dominantly used for answering RQs. Relying dominantly on qualitative methods also makes this study a \textit{qualitative study or qualitative in nature}.

To ensure the trustworthiness of the qualitative studies, Guba \cite{Guba1981} proposes credibility, transferablity, dependability, confirmability as criteria to be considered. Various techniques can be used to promote these criteria in qualitative studies. This study relied on techniques proposed by Shenton \cite{Shenton2004}.   

\subsection{Credibility of the Study}

Credibility denotes the alignment of study findings with the reality. To ensure the credibility, following techniques were used: \textit{The adaption of the well-established research method} this study used a survey to collect the data, qualitative coding and a clearly defined and analysis protocol developed by the InsighTD core group. Research methods that were used are also well accepted in software engineering field \cite{Singer2008,Seaman2008}.

\textit{Sampling} of subjects was done using the convenience sampling method, as presented in Section \ref{sec.DataCollection}. \textit{The honesty of information} was ensured by allowing potential participants to refuse to participate in the study, to quit the survey if they felt like doing so and by guaranteeing the anonymity of the participants. \textit{Frequent debriefing sessions:} during the analysis of data, researchers closely collaborated and all decisions and disagreements were solved with a consensus. \textit{Experience of the investigator:} The researchers who organized the InsighTD project and designed the survey were all experienced researchers. Junior researchers who participate in this project were under close supervision of their senior colleagues. \textit{Thorough description of the phenomenon under study:} this study focuses on TD and its manifestation in software development. TD is well recognized in software engineering literature, as it is presented in Section \ref{sec.RelatedWork}. 

The credibility of the study is also supported with the study organization. According to Santos et all. \cite{santos2019procedure}, organizing the study as a family of survey replications promotes the internal validity, which is the criteria used for credibility in quantitative studies \cite{Guba1981,Shenton2004}. Main reasons for this are the shared goals, RQs, study design, collaboration of independent research teams, shared raw data, and similar. Finally, the InsighTD survey was designed to ensure the credibility of the collected data. Fore example, Q13 allowed researchers to discard the answers based on whether the provided TD example could or could not be mapped to one of the TD types (see Section \ref{sec.Demographics.ExemplarProject}). 

\subsection{Transferability of the Study Findings}

Transferability denotes the application of study findings in different contexts or the generalizeability of findings. To promote the transferablitiy of the study findings, the technique of \textit{thorough description of the context} in which the study was executed was used. The context of this study is thoroughly described with a significant amount of details data in Section \ref{sec.Demographics} - Demographics and Section \ref{sec.InsTdProject} - InsighTD project. 

\subsection{Dependability on the Research Context}

Dependability denotes that the study repeated in the same context, with same participants, and same method would result with similar outcomes. To ensure the dependability of the study findings, a technique of \textit{Thorough description of the application of research methods} for data collection and analysis was used. Methodological descriptions provided in Section \ref{sec.ResearchApproach}, provide sufficient amount of data for interested researchers to replicate the study. 

\subsection{Confirmability of the Study Findings}

Confirmability denotes the objectivity of the findings, that is that the findings are based on the analysis of the collected data rather then on the researchers' beliefs. To ensure the confirmability of the study findings, the following techniques were used: \textit{Thorough methodological description}. Research approach adopted in this study is presented in Section \ref{sec.ResearchApproach}. This description enables the scrutiny of the study findings. \textit{Recognition of the study shortcomings}. Survey data can be incorrect due to: (a) lose of motivation to complete the survey or (b) when the participation in the study changes the behavior of the respondent. To mitigate these shortcomings, participants were familiarized with the study goal, topic, duration and that the participation in survey was anonymous.   

\section{Conclusion}
\label{sec.Conclusion}

The phenomenon of technical debt permeates the IT industry, and it is present in organizations and projects of various sizes. The large majority of developers is dealing with the consequences of injected TD at daily bases. It is evident that TD is a multifaceted phenomenon, and because of that, it is of utmost importance to collect rich empirical data about it. Current empirical investigations although important and valuable on its own have limitations regarding sample composition and size. Therefore, the objective of the InsighTD initiative is to coordinate research teams from different countries in order to jointly contribute to a single dataset about the TD phenomenon.  
So far, the InsighTD dataset includes 653 responses from 6 countries (Brazil, Chile, Columbia, Costa Rica, USA, and Serbia). 

Regarding the prevalence and familiarity of the TD concept, it seems that the concept is unknown to the almost third of participants (31\%). This percentage is relatively high, when compared to participants who had some practical experiences with TD identification and/or management (47\%). Furthermore, the analysis revealed that participants’ country has statistically significant impact on the TD concept prevalence. This strongly suggests that practitioners in some countries are more likely to be familiar and experienced with the concept, such as in Columbia and in the United States.  

Also, findings strongly suggest that a major cause of TD is \textit{deadline}, i.e. time pressure caused by short deadlines. And, interestingly the single most cited effect of TD is \textit{delivery delay}. The second most cited effect is \textit{low maintainability}. 
Based on this finding we can speculate that a \textit{TD vicious cycle}, i.e. a situation when existing debt incurs additional debt, can be caused and then fed by continuous time pressure.
Further data analysis showed that causes categorized as \textit{infrastructure} causes (environment setup, tools, and used technologies) will certainly lead to \textit{architectural debt}.

The analysis of effects did not reveal a single strong relation that stood out, like for causes, but rather several relations that describe expected effects of specific TD types. Some of them are well known, such as \textit{code debt} that leads to \textit{internal quality} issues, \textit{test debt} that leads to \textit{external quality} issues, and \textit{requirements debt} that leads to \textit{planning and management} issues. Alongside with those, \textit{defect debt} leads to \textit{people} related issues. It seems that postponing or avoiding corrections of known defects translates into growing dissatisfaction among developers as well as among users.

Families of surveys are rare in software engineering. InsighTD is the first family of surveys on technical debt in software engineering. 
As a research approach, family of surveys turned out to be very useful. It provided a methodological framework that allowed multiple replication teams to conduct research activities with a sufficient level of autonomy, but at the same time providing common grounds for aggregating data from multiple replications of the survey.

Regarding future work, further analysis of the data is planned in order to gain insight on undertaken practices to cope with TD. Furthermore, we expect to extend the InsighTD dataset with new replications. So far, we collected data from six out of twelve countries that joined the initiative. 



\bibliographystyle{model1-num-names}
\bibliography{mybibfile}

\end{document}